\documentclass{acm_proc_article-sp}
\usepackage[utf8]{inputenc}
\usepackage{times}
\usepackage{helvet}
\usepackage{courier}

\usepackage{xspace}
\usepackage{graphicx}
\usepackage{subfigure}
\usepackage{amsmath}
\usepackage{amssymb}
\usepackage{latexsym}
\usepackage{multirow}
\usepackage{array}
\usepackage{url}
\usepackage{paralist}
\usepackage{placeins}
\usepackage{cite}
\graphicspath{{figs/}}

\newcommand{\todo}[2][]{}

\newcommand{\comments}[1]{}

\newcommand{\eg}{e.\,g.,\xspace}
\newcommand{\ie}{i.\,e.,\xspace}

\newcommand{\conferator}{\textsc{Conferator}\xspace}

\makeatletter \@ifundefined{BibTeX}
   {\def\BibTeX{{\rmfamily B\kern-.05em%
    \textsc{i\kern-.025em b}\kern-.08em%
    T\kern-.1667em\lower.7ex\hbox{E}\kern-.125emX}}}{}
\makeatother

\newcommand{\furl}[1]{\footnote{\url{#1}}}

\def\sharedaffiliation{%
\newline
}

\usepackage[utf8]{inputenc}
\usepackage[lined,boxed,commentsnumbered]{algorithm2e}

\usepackage{tikz}
\usetikzlibrary{shapes,arrows,arrows,decorations.pathmorphing,backgrounds,positioning,fit,petri}

\begin{document}
\title{On the Predictability of Talk Attendance\\ at Academic Conferences}

\numberofauthors{1} %

\author{
\alignauthor
Christoph Scholz 
\quad
Jens Illig 
\quad
Martin Atzmueller  
\quad
Gerd Stumme\and
	   \sharedaffiliation
       \affaddr{Knowledge and Data Engineering Group, University of Kassel}\\
       \affaddr{Wilhelmsh\"{o}her Allee 73, D-34121 Kassel, Germany}\\
       \affaddr{\{scholz, illig, atzmueller, stumme\}@cs.uni-kassel.de}
}


\maketitle

\maketitle

\begin{abstract}
This paper focuses on the prediction of real-world talk attendances at academic conferences with respect to different influence factors. We study the predictability of talk attendances using real-world tracked face-to-face contacts. Furthermore, we investigate and discuss the predictive power of user interests extracted from the users' previous publications.
We
apply \textit{Hybrid Rooted PageRank}, a state-of-the-art unsupervised machine learning method that combines information from different sources. Using this method, we analyze and discuss the predictive power of contact and interest networks separately and in combination. We find that contact and similarity networks achieve comparable results, and that combinations of different networks can only to a limited extend help to improve the prediction quality. For our experiments, we analyze the predictability of talk attendance at the ACM Conference on Hypertext and Hypermedia 2011 collected using the conference management system \conferator.
\end{abstract} 

\section{Introduction}\label{sec:introduction}

Academic conferences facilitate scientific exchange, collaboration and innovation, \eg fostered by social contacts and interesting talks. A major task for every conference attendee is the selection of talks relevant to his research. Conference guidance systems such as \emph{Conference Navigator}~\cite{WBP:10} and \conferator\furl{http://www.conferator.org}~\cite{ubicon-2014a}, support this with the possibility of creating a personalized schedule. Picking talks manually, however, may become complex due to the large amount of available talks at a conference.
Furthermore, conversations with other attendees and changes in the conference schedule can influence the talk selection.


Recommendation components of conference guidance systems can support
their users by presenting suggestions of talks which the system determined as most interesting for the respective user. Then, such recommendations influence the decision \eg due to recommended talks which where otherwise not considered by the user.
Therefore, recommender systems should ideally always be evaluated in an online scenario, where influence is part of the evaluation. 


In this paper, we focus on the predictability of real talk attendances,
\ie we try to find models imitating the actual decision process without recommendation influence. Due to the low availability of online recommender evaluation, it is reasonable to evaluate recommender systems on a prediction setting. This is partially valid since good predictions are also good recommendations to the extent that the user does not repent the predicted decisions. 
For our evaluation, we use real-world talk attendance data which was collected using \conferator.
\conferator applies active RFID technology developed by the SocioPatterns consortium\furl{http://www.sociopatterns.org} for the localization as well as for the measurement of face-to-face contacts between researchers during the conference, \eg during the coffee breaks. Based on such RFID data and
collected content information of scientific papers, we investigate the potential of social contact information and content-similarity for predicting real-world talk attendance decisions. Especially, we analyze the potential of combining different information sources for improving the overall prediction quality.

Our contribution can be summarized as follows:
\begin{enumerate}
\item We present the first study about the predictability of visited talks at academic conferences on real world data.
\item We analyze different influence factors concerning the predictability of talks at academic conferences. In particular, we study the influence of face-to-face contacts and user interest on the talk attendance decision.
\item We consider and adapt state-of-the-art unsupervised link prediction methods for the talk prediction problem, focusing on the \emph{rooted PageRank}~\cite{Kleinberg2003} and the \emph{Hybrid Rooted PageRank}~\cite{SABCS:13} algorithms.
\item We present an in-depth analysis of talk attendance predictability using different performance metrics and investigate the influence of different interaction networks, \eg derived from social contact and content information, for this task. 
\end{enumerate}

The rest of this paper is structured as follows: Section~\ref{sec:related} discusses related work. In Section~\ref{sec:conferator} we describe the framework that we used to collect our data. Section~\ref{sec:dataset} gives a detailed overview of the collected dataset. In Section~\ref{sec:algorithms}, we discuss the algorithms used for the prediction task. After that, Section~\ref{sec:evaluation} presents a detailed evaluation using the dataset collected at ACM Hypertext 2011. Finally, Section~\ref{sec:conclusion} summarizes our results.

\section{Related Work}\label{sec:related}

In this section, we discuss related work concerning the talk prediction problem at academic conferences. We start with relevant work about the analysis of human contact pattern at conferences and then discuss work about talk recommendation.

\subsection{Analysis of Human Contact Patterns and Link Prediction}
 
 The analysis of offline social networks, focusing on human contacts, has been largely neglected. In this context,  Eagle et al. \cite{eagle2009} and
 Zhoe et al.~\cite{Hui2005} presented an analysis of proximity information collected by devices based on Bluetooth communication, similar to Xu et al.~\cite{XCWCZYWZ:11}, who also related this to online social networks.
 However, in all these experiments it was not possible to detect reliable face-to-face contacts. The SocioPatterns collaboration developed
an infrastructure that detects close-range and face-to-face proximity (1-1.5 meters)
of individuals wearing proximity tags with a temporal resolution of 20 seconds~\cite{Cattuto:2010}. 
Due to the fact that the human body blocks RFID signals this allows the detection of face-to-face contacts between persons. One of the first experiments using this kind of proximity tags was done by by Cattuto and colleagues in~\cite{ALANI09}. They presented an application that combines online and offline data from conference attendees. In \cite{Zuo:2012} the authors also studied the influence between
offline and online properties using a mobile social application in the context of academic conferences. Barrat et al.~\cite{Barrat10} compared the attendees' contact patterns with their research seniority,
their co-authorship and their activity in social web platforms. The SocioPatterns sensing infrastructure was also deployed in other environments in order
to study the dynamics of human contacts, such as healthcare environments~\cite{Isella:2011}, schools~\cite{Stehl2011} and museums \cite{DBLP:journals/corr/abs-1006-1260}.
Atzmueller et al.~\cite{ADHMS:12} described the dynamics of community structures and roles at conferences, extending the analysis of interactions and dynamics, and the connection between research interests, roles and academic jobs of conference attendees~\cite{MSAS:12}.

Link prediction, as defined by Liben-Nowell and Kleinberg in \cite{Kleinberg2003}, is strongly related to talk prediction. In \cite{Kleinberg2003} the authors did a first comprehensive analysis by analyzing the predictability of unsupervised machine learning methods. Scholz et al. analysed the predictability of face-to-face contacts at academic conferences. In \cite{SABCS:13} the authors presented an unsupervised link prediction method that combines information of different networks. 

\subsection{Talk Recommendation and Prediction}
To the best of our knowledge, predictability of scientific talk attendance has not yet been
investigated w.r.t the true physical attendance of conference talks.
Talk recommendation is a specific instance of the general recommendation task.
Published work about talk recommendation systems evaluated the recommendation algorithms with respect to their ability to reconstruct the remaining part of a partly given user's attendance plan entered into some conference management system.

For recommender systems, we typically distinguish between content-based and
collaborative-filtering approaches~\cite{AT:05}. Content-based recommenders make use of
properties of the recommended items, while collaborative filtering methods utilize common item ratings of users. For talk recommendation, items are talks, while author, title, and abstract are content-properties.
%

Minkov et al.~\cite{minkov2010} as well as Pham et al.~\cite{pham2012} simulated talk attendances and collected explicit user feedback in form of questionnaires about the generated recommendations. Based on this feedback they evaluate their algorithms.
%
Both evaluation schemes have their drawbacks. Using
attendance plans as a gold-standard is not absolutely correct, because it is
unclear whether the user actually attended the talks. Instead, plans may be non-final
or simply be a collection of bookmarks used as a reminder for later attendance
decisions. Furthermore, the plans may be incomplete in terms of not covering
all time slots of the conference.
Using questionnaires, the user is usually only asked once or few times about her
satisfaction with the recommendations. Therefore, questionnaires give only a very
rough measure about the overall recommendation quality, contrasting the evaluation for the prediction task where there are quantitative evaluations based on the correctness of each single prediction. In questionnaires, users
can also rate the recommendation quality high if the system recommended a talk which
the user did not attend but still found interesting. The other way round, attended
talks may have bored the user and could therefore be bad recommendations although
being perfect predictions. Questionnaires therefore measure a slightly different property than measured when evaluating in a prediction setting.

Minkov et al.~\cite{minkov2010} trained a RankSVM \cite{joachimsRankSvm} classifier by supervision from a training part of their user feedback and evaluated on a test set. They augmented their 
content-based approach with a collaborative aspect using a modified RankSVM optimization problem which integrates dimensionality
reduction and optimizes the dimensionality reduction parameters across users.
In contrast to this work, our work uses a more explicit usage of social networks and
focuses on unsupervised or weakly supervised approaches since we expect little
knowledge about the talk attendance preferences of most conference visitors.

Pham et al.~\cite{pham2012}, as well as Lee and Brusilovsky~\cite{lee2012} applied collaborative filtering for the
recommendation of talks.
For each user, \cite{pham2012} calculate sets of the 5 most similar users
according to either commonly bookmarked talks in \emph{Conference Navigator} or
common co-authors. They furthermore use content-boosted collaborative filtering.
User-similarity is then not solely calculated from known shared
bookmarks or known co-authored publications (with a weight of 1), but also from
relations to other users which have bookmarks of talks or co-authored papers similar to those
the target user is related to (using the maximum cosine similarity to any of the
other users bookmarks). Results were further enhanced by reweighting scores based
on numbers of common co-authors and co-authored papers but precision was reported
to be low with a maximum of $21\%$.
Lee and Brusilovsky~\cite{lee2012} rely on bookmarks obtained from the conference management system
\emph{Conference Navigator 2.0} \cite{WBP:10}. This approach also uses boosted collaborative filtering. But instead of adding
content similarity to the similarity based on bookmarks (or planned attendances),
\cite{lee2012} calculate most similar users based on a weighted average of Jaccard coefficients on common co-authors and commonly referenced publications. 
They evaluated recommendation quality based on \emph{conference
simulations} where feedback from six \emph{evaluators} was retrieved using
evaluation forms. They provided two textual example statements but no quantitative
evaluation.


The difference between our work and existing literature is that we present the first analysis of the predictability of visited talks at conferences using real world data. Furthermore, we study the influence of face-to-face contacts and user interest concerning the talk prediction problem. In particular, we consider combinations of different knowledge sources given as social interaction networks.

\section{Conferator -- A Social Conference Management System}\label{sec:conferator}
In the following section, we first outline the active RFID technology used by the \conferator system. Next, we introduce the \conferator and its functionality.

\subsection{RFID-Setup}\label{sec:dataset:setup}
\begin{figure}[htb]
  \begin{minipage}{.45\columnwidth}
    \begin{center}
     \includegraphics[height=50pt]{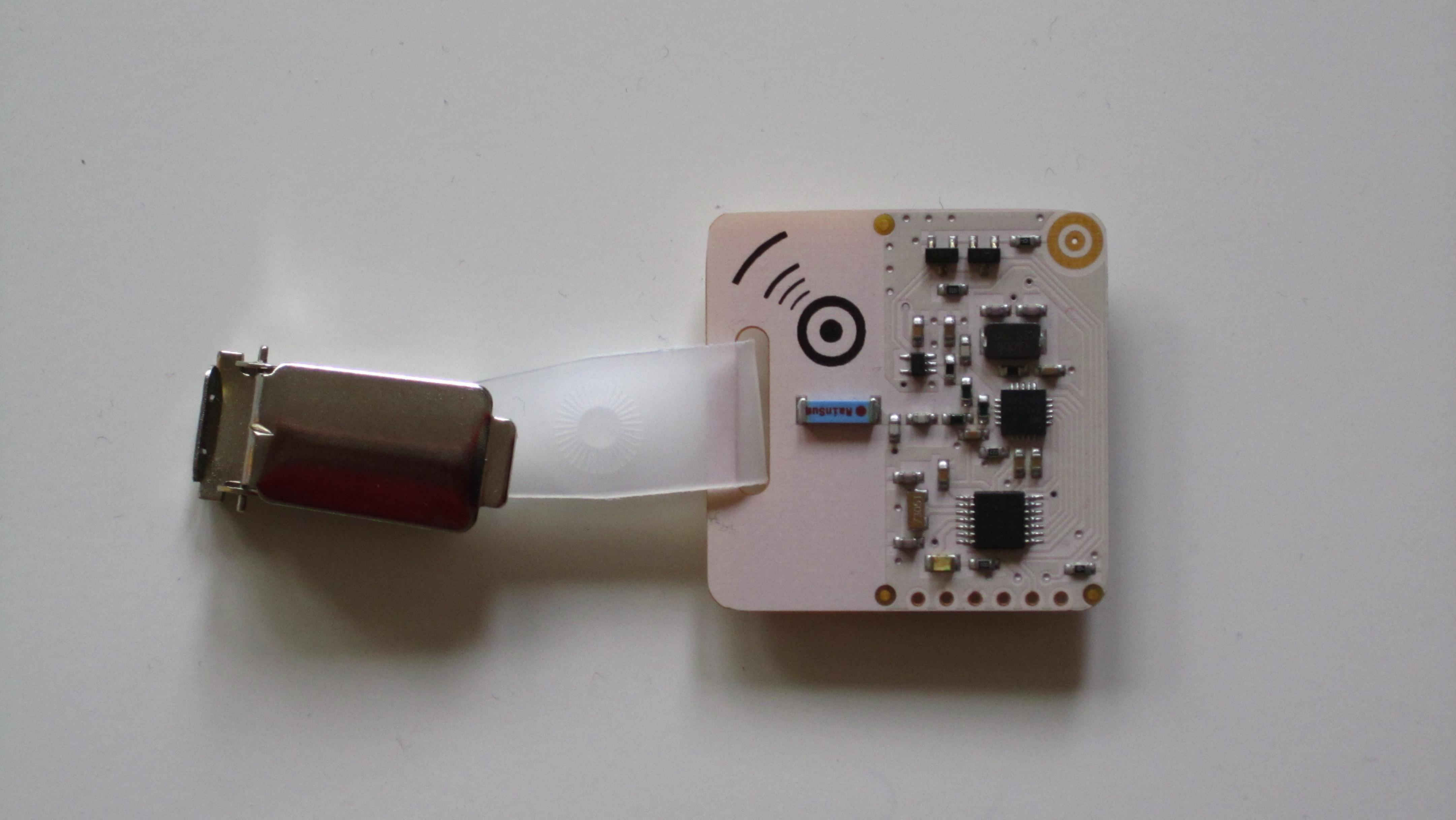}
    \end{center}
 \end{minipage}
  \hfill
  \begin{minipage}{0.45\columnwidth}
    \begin{center}
      \includegraphics[height=50pt]{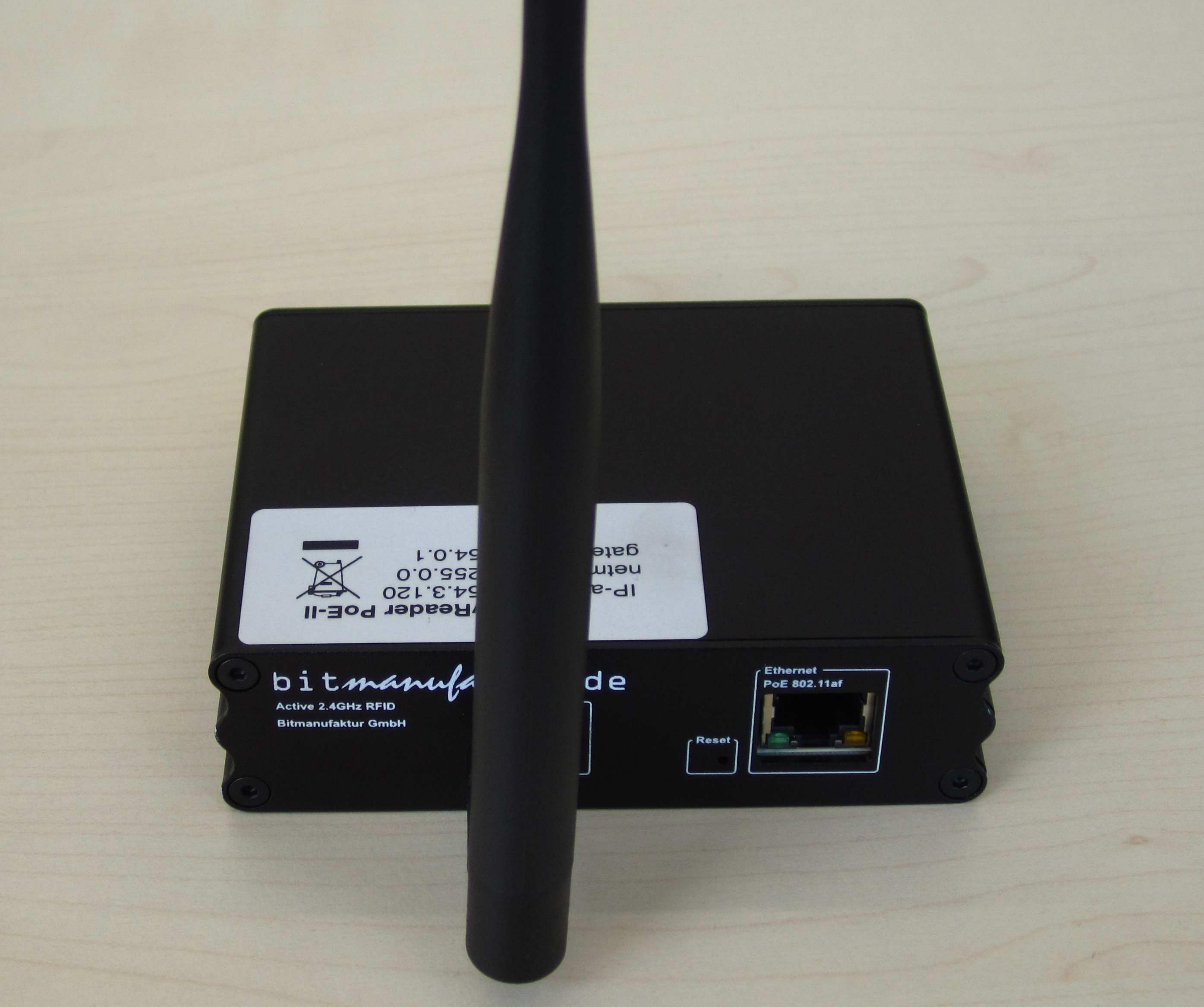}
    \end{center}
  \end{minipage}
  \caption{Proximity Tag (left) and RFID Reader (right)}
  \label{fig:tag_reader}
\end{figure}

At the Hypertext 2011 conference we asked each participant to wear an active RFID tag (see Figure \ref{fig:tag_reader}). One decisive factor of these active RFID tags is the possibility to detect other active RFID tags within a range of up to 1.5 meters, which allows us to create human face to face contact networks. We call these active RFID tags \emph{proximity tags} in the following. Each proximity tag sends out two types of RFID-signals, proximity signals and tracking signals. A proximity signal is used for contact sensing, which is achieved by using signals with very low radio power levels ~\cite{CIROHIGHRES08}. The proximity tag sends out tracking signals in four different signals strengths (-18dbm, -12dbm, -6dbm, 0dbm) to RFID readers (see Figure \ref{fig:tag_reader}) placed at fixed positions in the conference area. These tracking signals are used to transmit proximity information to a central server and for determining the position of each conference participant ~\cite{SDAHS:11}\cite{WCI:07}. Depending on the signal strength the range of a tracking signal inside a building is up to 25 meters. Each signal contains the signals strength and ID of the reporting tag and the IDs of all RFID tags in proximity. For more information about the proximity tags we refer to Barrat et al ~\cite{CIROHIGHRES08} and the OpenBeacon website.\footnote{\url{http://www.openbeacon.org}}      

\subsection{Conferator}
\begin{figure}[hbt]
     \begin{center}
      \includegraphics[height=170pt]{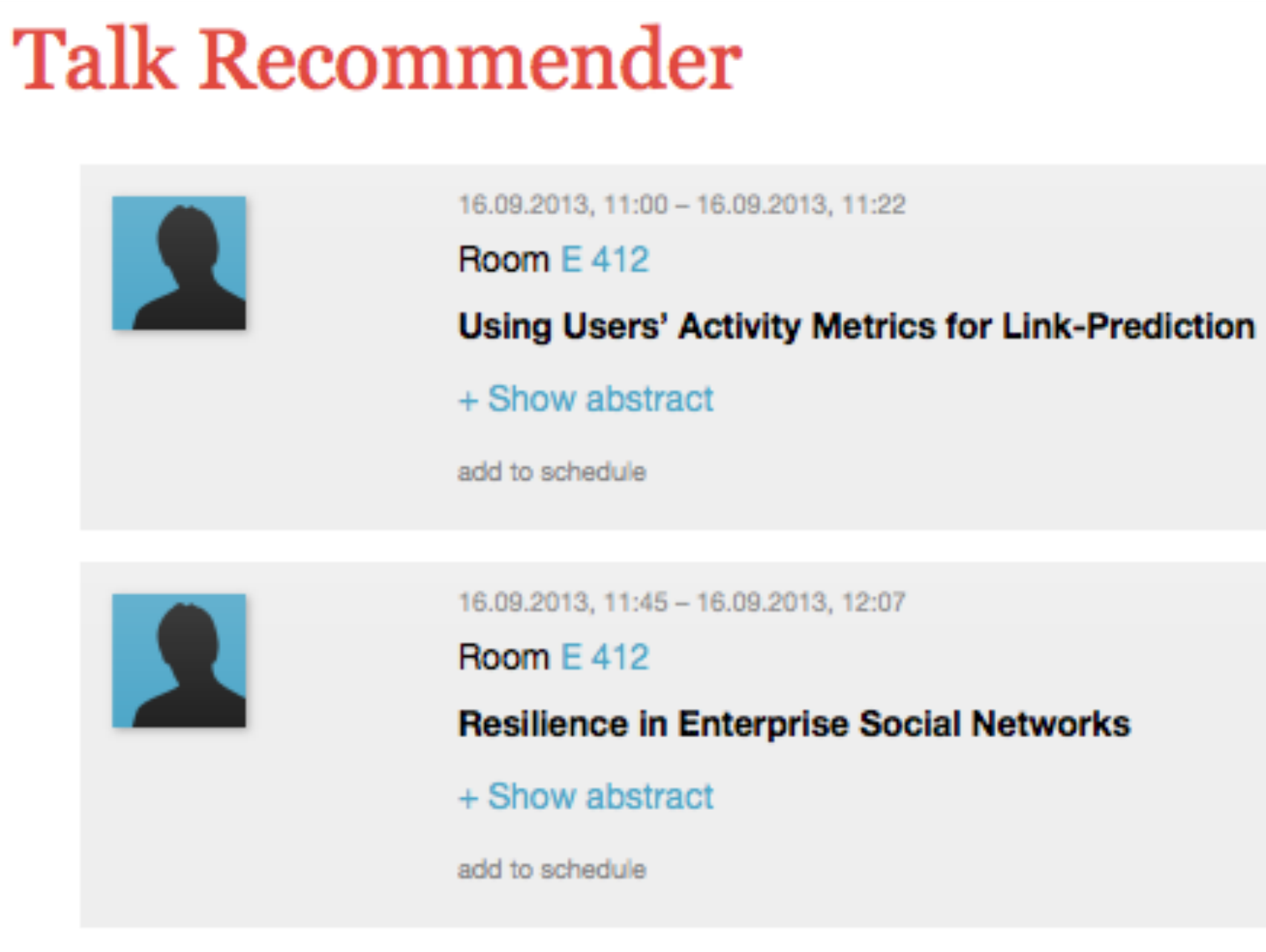}
      \end{center}
      \caption{Talk recommender in the social conference management system \conferator.}
\label{fig:talkrecommender}
\end{figure}

The \conferator system~\cite{ubicon-2014a} is a social and ubiquitous conference guidance system. It allows conference participants to manage their conference schedule. Furthermore, the \conferator  supports social interaction at a conference. For example, it is possible for conference participants to recall their own contacts
 or to browse through other conference attendees' user profiles. Furthermore, the \conferator presents personalized suggestions for interesting talks (see Figure~\ref{fig:talkrecommender}). We note here that the recommendation component presented in Figure~\ref{fig:talkrecommender} was not a part of the \conferator at HT 2011. Here the talk recommendations were provided by the Conference Navigator~\cite{WBP:10}.   
\conferator has successfully been deployed at several events, e.g., the LWA
2010\footnote{\url{http://www.kde.cs.uni-kassel.de/conf/lwa10/}}
, LWA 2011\footnote{\url{http://lwa2011.cs.uni-magdeburg.de/}} and LWA 2012\footnote{\url{http://lwa2012.cs.tu-dortmund.de/}} conferences,
the Hypertext 2011\footnote{\url{http://www.ht2011.org/}} conference, the INFORMATIK 2013\footnote{\url{http://informatik2013.de/}} conference, and a technology day of the Venus\footnote{\url{http://www.iteg.uni-kassel.de/}} project.   
In this paper, we focus on data collected at the Hypertext 2011.

\section{Datasets}\label{sec:dataset}
In the following section we introduce the dataset collected at the $22^{nd}$ ACM Conference on Hypertext and Hypermedia 2011 (HT 2011) in Eindhoven. We present statistics characterizing key properties of the applied data.

\subsection{Face-to-Face Contact Data}

\begin{figure}[hbt]
\subfigure[]{
\begin{minipage}{.97\columnwidth}
      \begin{center}
      \includegraphics[width=0.82\columnwidth]{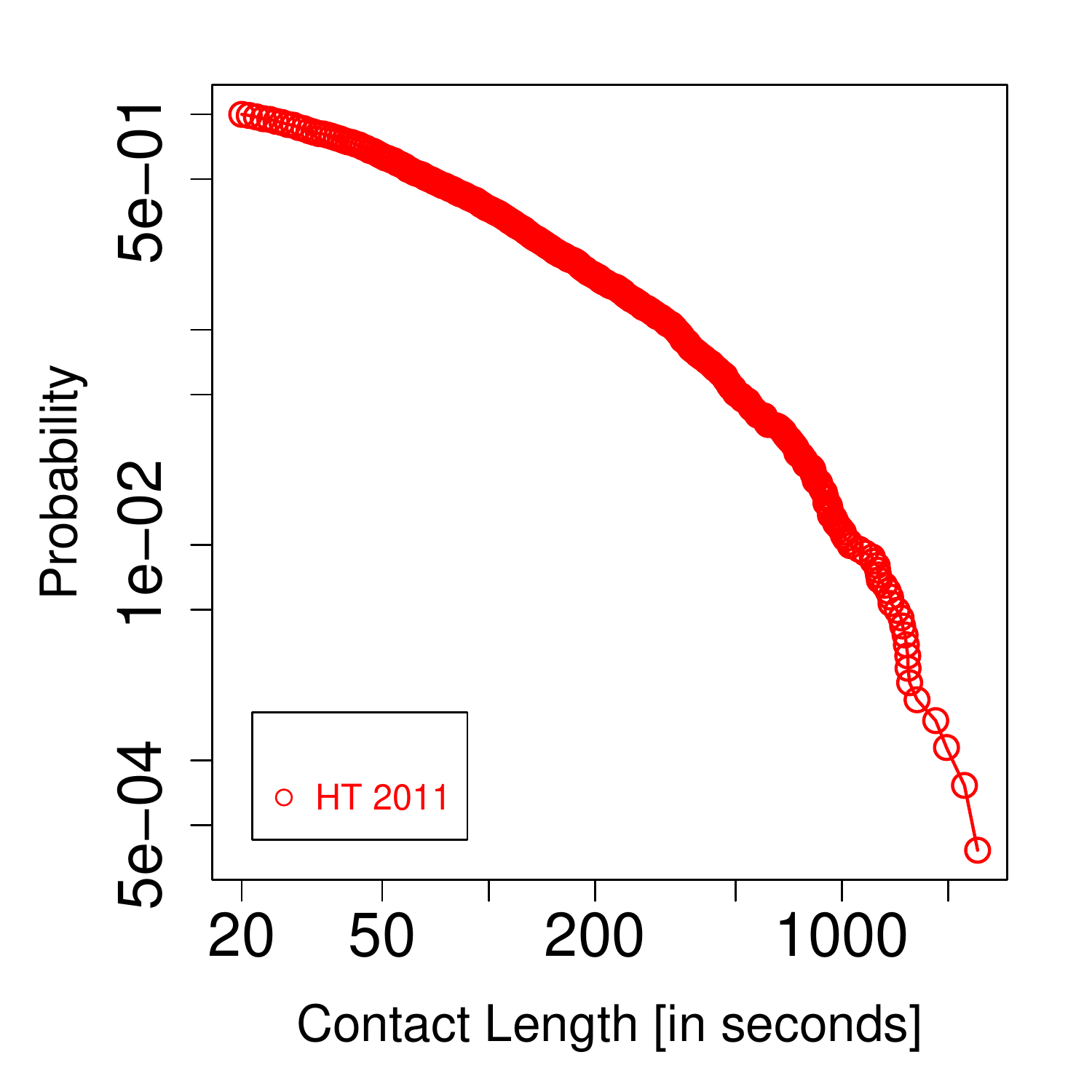}
       \end{center}
\end{minipage}
\label{fig:CL_ALLCONTACTS}
}
\hfill
\subfigure[]{
\begin{minipage}{.97\columnwidth}
      \begin{center}
      \includegraphics[width=0.8\columnwidth]{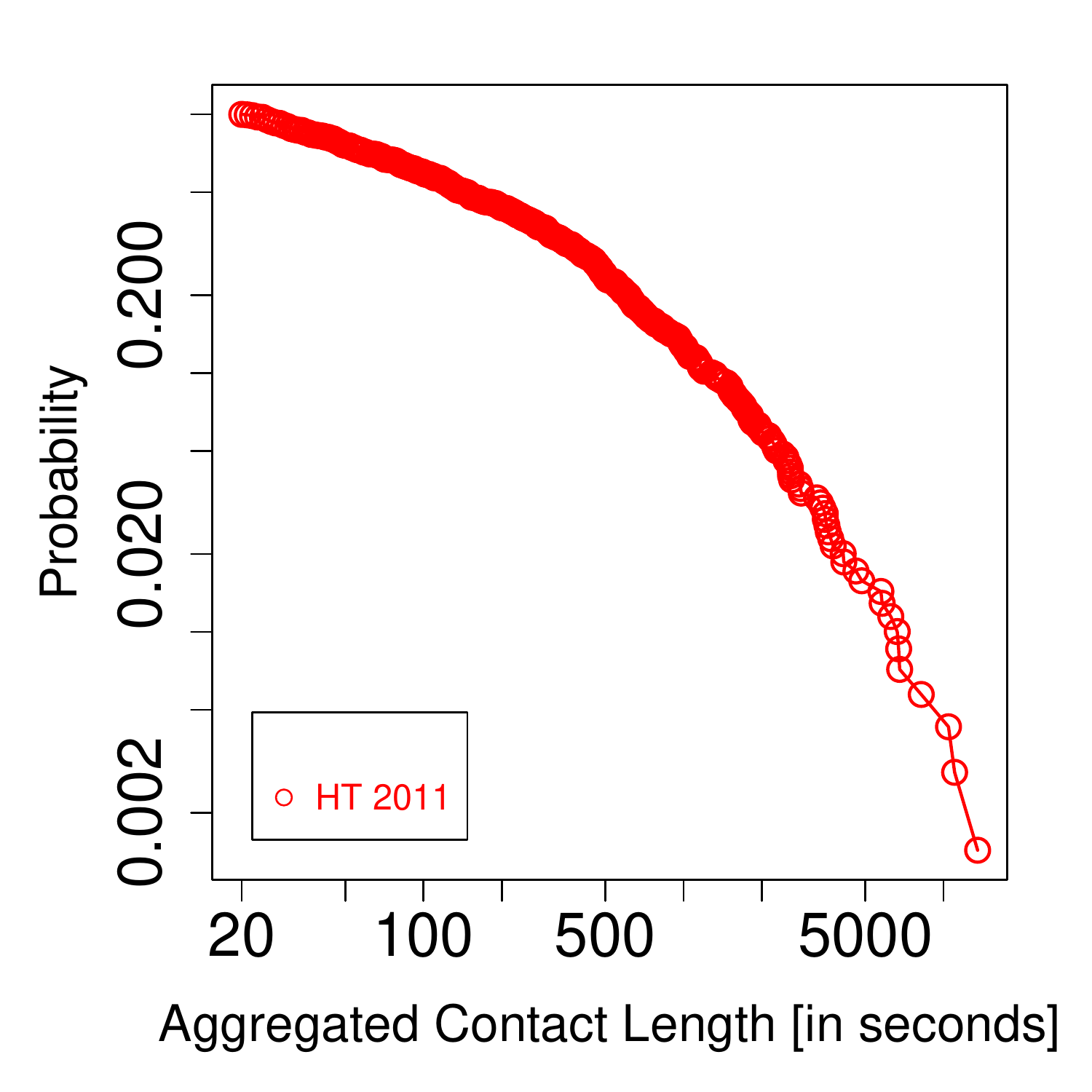}
       \end{center}
\end{minipage}
\label{fig:CL_AGGCONTACTS}
}
 \caption{Cumulated contact length distribution of all face-to-face contacts (a) and all aggregated face-to-face contact at the HT-2011 conference. The $x$-axis displays the minimum length of a contact in seconds, the $y$-axis the number of contacts having at least this contact length, respectively. The axes are scaled logarithmically.}
\label{fig:ProbRecContact}
\end{figure}

Table~\ref{tab:datasetstatistics} provides a summary on the characteristics of the collected face-to-face proximity dataset. As already observed before~\cite{Cattuto:2010,DBLP:journals/corr/abs-1006-1260,MSAS:12}, the distributions of all contacts and all aggregated face-to-face contacts lengths between conference participants are
heavy-tailed (see Figures \ref{fig:CL_ALLCONTACTS} and \ref{fig:CL_AGGCONTACTS}). 
 More than half of all aggregated face-to-face contacts are shorter than 200 seconds and the average contact duration is less than one minute. However, very long contacts are also observed.
The diameter, average degree,
and average path length of $G$ are similar to the results presented in
\cite{DBLP:journals/corr/abs-1006-1260,ADHMS:12}.
For more details on the applied dataset, we refer to, \eg~\cite{SABCS:13} and ~\cite{MSAS:12} .

\begin{table}[htb]
    \centering
    {
    \begin{tabular}{|l|c|}
    \hline
    & \textbf{HT 2011} \\
    \hline\hline
    \textit{\#days} & $3$\\
    \hline
    \textit{$|V|$} & $68$\\
    \hline
    \textit{$|E|$} & $698$\\
    \hline
    \textit{Avg.Deg.($G$)} & $20.53$\\
    \hline
    \textit{APL ($G$)} & $1.76$\\
    \hline
    \textit{d ($G$)} & $4$ \\
    \hline
    \textit{AACD} & $529$ \\
    \hline
    \end{tabular}
    }
\caption{Collected dataset at HT 2011. Here $d$ is the diameter, AACD the average aggregated contact-duration (in seconds) and \textit{APL} the average path length.}
       \label{tab:datasetstatistics}
\end{table}

\subsection{Talk Attendance Data}

For our analysis, we focus on the parallel talks at HT 2011. Overall, 14 parallel talks took place in two rooms. For our prediction analysis it is essential to determine whether or not a participant attended a talk. Therefore, we installed one RFID reader in each conference room. As described in Section~\ref{sec:dataset:setup}, a proximity tag sends out tracking signals that we used to determine the current position of each conference participant at room level basis. For the determination of talk attendance we used the following localization strategy: Since the walls of each conference room (where the talks took place) were very thick and hence tracking signals (sent out by the proximity tags) could only be detected within one conference room. This means, that if we detect a tracking signal of a conference participant in a conference room, then we know that that this participant must be in this room. Overall, we observed 359 visited talks from 53 conference participants.

\subsection{Full-Text Data}

\begin{figure}[hbt]
     \begin{center}
      \includegraphics[width=0.7\columnwidth]{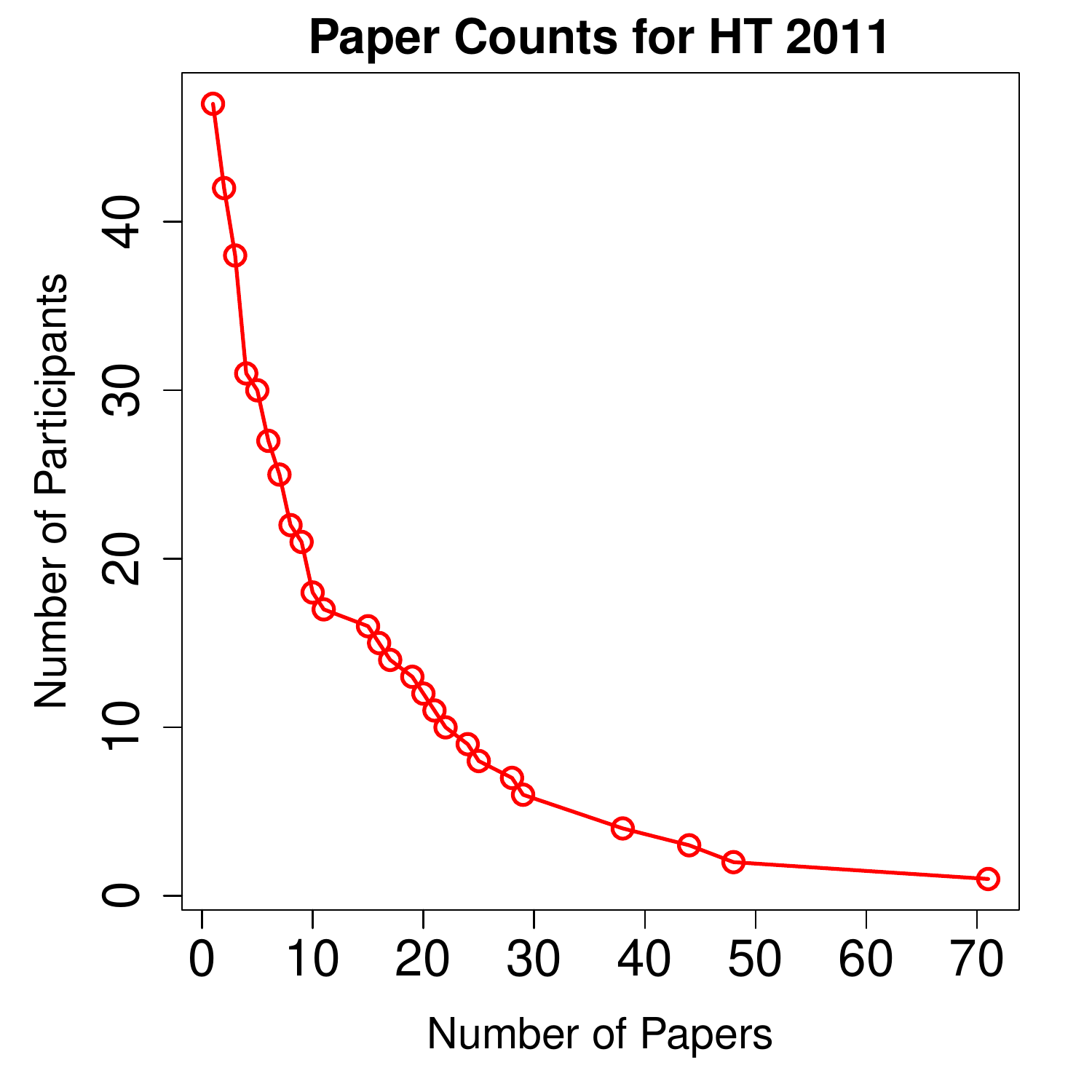}
      \end{center}
      \caption{Distribution of the number of papers for each conference participant. The $y$-axis displays the number of participants having at least the number of papers that are defined by the $x$-axis.}
\label{fig:paperdistribution}
\end{figure}

For our prediction task, we also consider the content of all papers. For each conference participant, we therefore crawled all papers that are listed in DBLP since 2006. In total, we crawled 707 papers. With the full-text data we created bag-of-words models representing the paper profiles for each participant. For the participants' bag-of-words-model construction, we used the Porter Stemmer algorithm~\cite{PorterStemmerAlgorithm} and removed all stop words. Figure~\ref{fig:paperdistribution} displays the cumulative number of papers for each conference participant.

\section{Algorithms}\label{sec:algorithms}
In this section, we describe the algorithms used for the prediction of talks at academic conferences. Focusing on unsupervised methods, we use the \emph{Hybrid Rooted PageRank} algorithm, an extension of the \emph{rooted PageRank} algorithm, for prediction.

\subsection{Rooted PageRank}
The \emph{rooted PageRank} predictor (RPR) \cite{Kleinberg2003} is an adaption of the \emph{PageRank} algorithm \cite{BrinP98} for the link prediction task. The \textit{rooted PageRank} predictor score between participants $r$ and $y$ is defined by the stationary probability distribution of participant $y$ under the following random walk \cite{Kleinberg2003}:
\begin{compactitem}
\item With probability $\alpha$, jump to $r$.
\item With probability $1-\alpha$, jump to a random neighbor of the current node.
\end{compactitem}
For the \textit{weighted rooted PageRank (WRPR)} predictor, the random walk selects a random neighbor $n$ of the currentnode $c$ with probability $\frac{w(c,n)}{\sum\limits_{c \rightarrow d}{w(c,d)}}$, where $w(c,d)$ is the weight of the edge $(c,d)$.

\subsection{The Hybrid Rooted PageRank Method}\label{sec:algorithms:hrpr}
In this section, we describe the \textit{Hybrid Rooted PageRank} algorithm, first presented in~\cite{SABCS:13}. This algorithm is an unsupervised machine learning method and extends the \textit{rooted PageRank} algorithm. The \textit{Hybrid Rooted PageRank} algorithm combines the information of different networks. To do so, the \textit{Hybrid Rooted PageRank} computes the stationary distribution of nodes under the random walk described
in Algorithm~\ref{alg:hybridrandomwalk}. In each step, the walk selects a given network with respect to a given probability distribution. From the current node $c$ a link in this network is then selected to a random neighbor $n$ of node $c$ with probability $\frac{w(c,n)}{\sum\limits_{c \rightarrow d}{w(c,d)}}$, where $w(c,d)$ is the weight of the edge $(c,d)$. If no link exists in the chosen network
(i.e., if the node is isolated),
then the algorithm jumps back to the root node. In this way, one can integrate different networks for prediction of links.

\IncMargin{2em}
\begin{algorithm}
\SetKwData{Left}{left}\SetKwData{This}{this}\SetKwData{Up}{up}
  \SetKwFunction{Union}{Union}\SetKwFunction{FindCompress}{FindCompress}
  \SetKwInOut{Input}{Input}\SetKwInOut{Output}{Output}
 \SetAlgoLined
 \LinesNumbered
 \DontPrintSemicolon
  \Input{Networks $N=\{N_1,\ldots,N_n\}$, Network-Probabilities $P=\{p_1,\ldots,p_n\}$, Probability $\alpha$, Root node $r$}
 \Output{Stationary distribution weight of node $v$ under the following random walk:}
\BlankLine
 With probability $\alpha$ jump to root node $r$.\;
 With probability $1-\alpha$:\;
 \Indp Choose Network $N_i\in N$ with respect to probability distribution $P$.\;
 \lIf {There exist no outgoing edges}\; {\Indp Jump to root node $r$\;
 \Indm \lElse{From the current node $c$ jump to a neighbor $n$ selected with a probability $\frac{w(c,n)}{\sum\limits_{c \rightarrow d}{w(c,d)}}$, \ie  proportional to the weight $w(c,n)$ of the edge $(c,n)$. }
 }
   \caption{\textit{Hybrid Rooted Random Walk}}
   \label{alg:hybridrandomwalk}
 \end{algorithm}\DecMargin{2em}

 Assume we want to determine the \textit{Hybrid Rooted PageRank} predictor score for participants $r$ and $x$. In this case, we use participant $r$ as root node and execute the algorithm. As a result, the algorithm computes the stationary probability distribution of all nodes. The predictor score between participant $r$ and $x$ is then given by the stationary probability distribution of participant $x$.

\section{Evaluation}\label{sec:evaluation}
In this section, we analyze the predictability of talk attendance at academic conferences. Specifically we study the influence of face-to-face contacts and user interests on this prediction problem. Furthermore we consider combinations of different knowledge sources given as social interaction networks. We start with an explanation of the used evaluation methods, before we present and discuss the predictability results. 

\subsection{Evaluation Method}
In this section we define and discuss the measures that we calculate for evaluating the impact of
various examined influence factors on real world talk attendance decisions. We use
two measures: \emph{Accuracy} and \emph{area under the curve} of the
\emph{receiver operating characteristic}.

\subsubsection{Accuracy}
Accuracy (ACC) is widely used and simply refers to the fraction of correct
decisions divided by the total amount of decisions.

The problem of predicting attended talks, which we cover in this work, is
distributed over time slots. Naturally, for every person only one talk can be
attended at each time slot. Thus, for each time slot, one decision \textit{has} to be made for every conference
participant that attends one of the
parallel talks. Applied to our talk prediction setting, accuracy can be
interpreted as the maximum likelihood probability estimate for a talk
recommender system to correctly predict the next attended talk.

\subsubsection{Area Under the Curve of the Receiver Operating Characteristic}
We further use the area under the receiver operating characteristic (here simply
abbreviated as AUC)~\cite{Hanley1982}. The
receiver operating characteristic (ROC) is given by a plot, which is defined as follows. For each n, a point is added to the curve, based on the top $n$ decisions of the algorithm ranked by relevance. The $x$-coordinate of the point is the false-positive rate of these $n$ decisions, and its $y$-coordinate is the true-positive rate.

For our talk prediction task, we use AUC to evaluate a conference-global ranking. For each pair of parallel talks $t_1$ and $t_2$ and talk attendee $p$ we calculate two predictor scores, one for $t_1$ and one for $t_2$. This results in a ranking containing all positive and negative decisions for predicting all talks. AUC evaluation rewards a predictor's ability to rank correct decisions before wrong decisions according to the ground truth. An ideal
predictor ranks all correct decisions above all wrong
predictions and achieves thus an AUC score of $1.0$, while a purely random predictor
achieves a score of $0.5$.

\subsection{Predictability of Talk Attendance}

In this subsection we study and discuss the predictability of talk attendance at academic conferences. We start with first statistics concerning the talk and session attendance behavior at the HT 2011 conference.

\subsubsection{Talk and Session Attendance Statistics}
  In Table~\ref{tbl:SessionStatistics} we present first statistics about the talk and session attendance behavior at HT 2011 for the parallel talks. Overall, the $53$ conference participants attended $194$ sessions. We observe that most of the participants did not change a session during the HT 2011. At this conference, only in $7 \% (\frac{14}{194})$  of all cases, the corresponding participant changed the session. In $69 \% (\frac{134}{194})$ of all cases the participants visited all talks of the session.

\begin{table}[thb]
    \centering
    {
    \renewcommand{\arraystretch}{1.04}
    \begin{tabular}{|l|c|}
    \hline
    \textbf{$\#$ Sessions} & 194\\\hline
    \textbf{$\#$ Visited All Talks in Session} & 134\\\hline
    \textbf{$\#$ Changed Session} & 14\\\hline
    \textbf{$\#$ Visited exactly 2 talks of Session} & 13\\\hline
    \textbf{$\#$ Visited exactly 1 talk of Session} & 33\\
    \hline
    \end{tabular}
    }
     \caption{Statistics about talk and session attendance behavior at the HT 2011 conference.}
     \label{tbl:SessionStatistics}
\end{table}

\subsubsection{Influence Factors of Talk Attendance Using Face-To-Face Contact Networks}
\begin{figure}[hbt]
     \begin{center}
      \includegraphics[width=1.02\columnwidth]{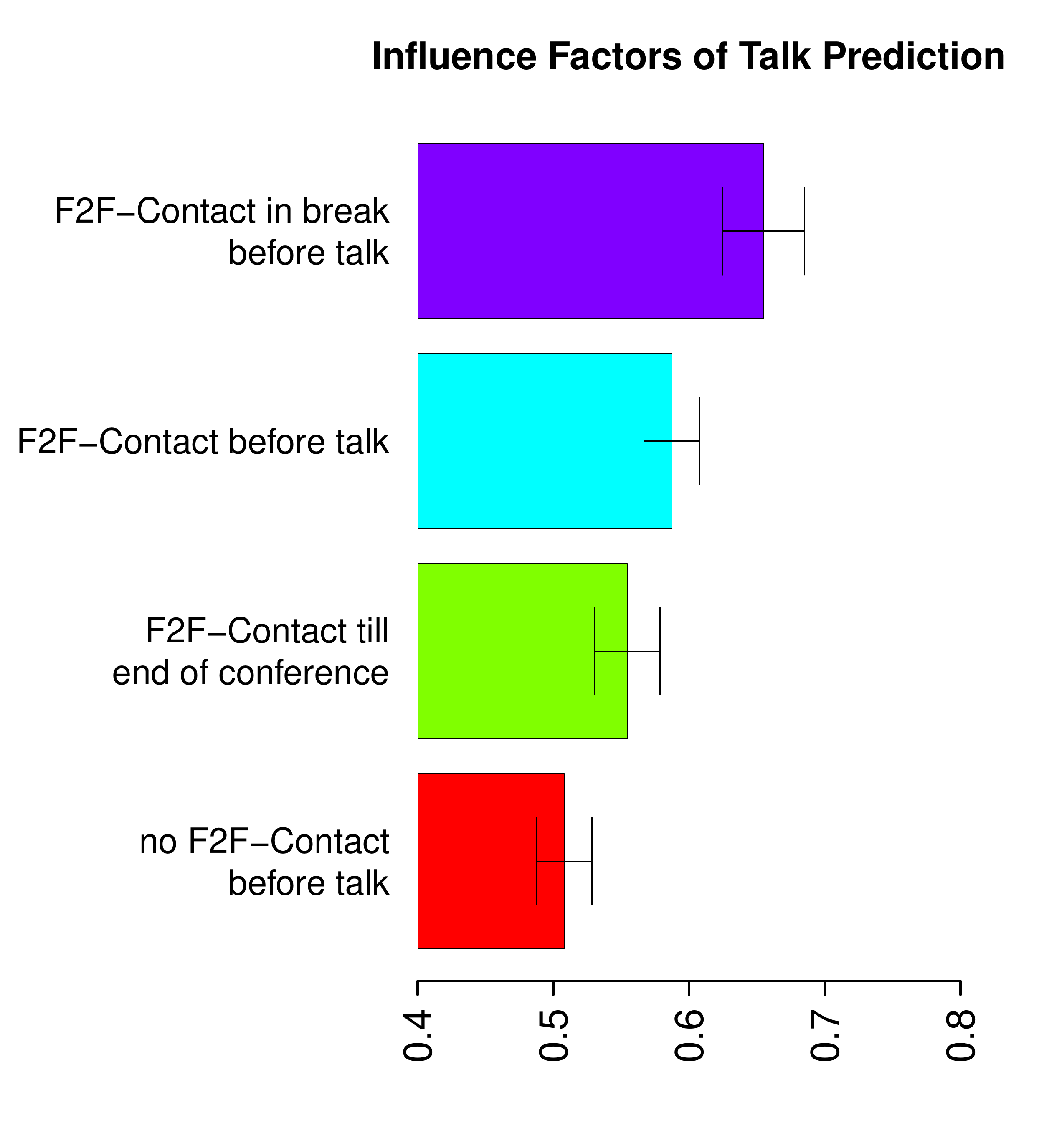}
      \end{center}
      \caption{Analysis of influence factors concerning the prediction of talks at academic conferences. Here we plot the probability that two participants visit the same talk, given that 1. there is a face-to-face contact in the coffee break before the next talk  is going to start, 2. there exists a face-to-face contact before the next talk is going to start 3. there exists a face-to-face contact till the end of the conference, 4. there exists no face-to-face contact.}
\label{fig:influencefactor_f2f}
\end{figure}

 In this section, we study the influence of face-to-face contacts during a conference on the attendance of talks. Especially we analyze the probability that two participants attended the same talk, based on the current face-to-face contact behavior between these two participants. In the following, we apply a t-test for determining the significance of our observations. We therefore will also plot the $95 \%$ confidence intervals of the results.
 First, we assume that there exists no face-to-face contact between two conference participants until the start of talk $t$. In Figure~\ref{fig:influencefactor_f2f}, we observe that the probability is nearly random (\ie probability is $50.8 \%$) that these two participants visit the same talk $t$, if there exists no prior face-to-face contact. In addition, we analyze the probability that two participants visit the same talk, when there exists a face-to-face contact till the end of the conference. (Note that this information could not be used for our prediction task, because it used future information.) We observe that the probability here is slightly increased (probability is $55.5 \%$) to attend the same talk, if there will exist a face-to-face contact till the end of the conference. It is interesting to see that the probability is $58.74\%$, if already a prior face-to-face contact exists, before the talk starts. This result highlights the influence of face-to-face contacts on the talk attendance. Furthermore, we analyzed  whether a face-to-face contact during the coffee break will influence the probability to attend the same talk of the next session. In Figure~\ref{fig:influencefactor_f2f}, we see that that the probability is $65.5 \%$ to attend the same talk of the next session, if there exists a face-to-face contact in the coffee-break before the session.   
 
\begin{figure}[hbt]
     \begin{center}
      \includegraphics[width=0.85\columnwidth]{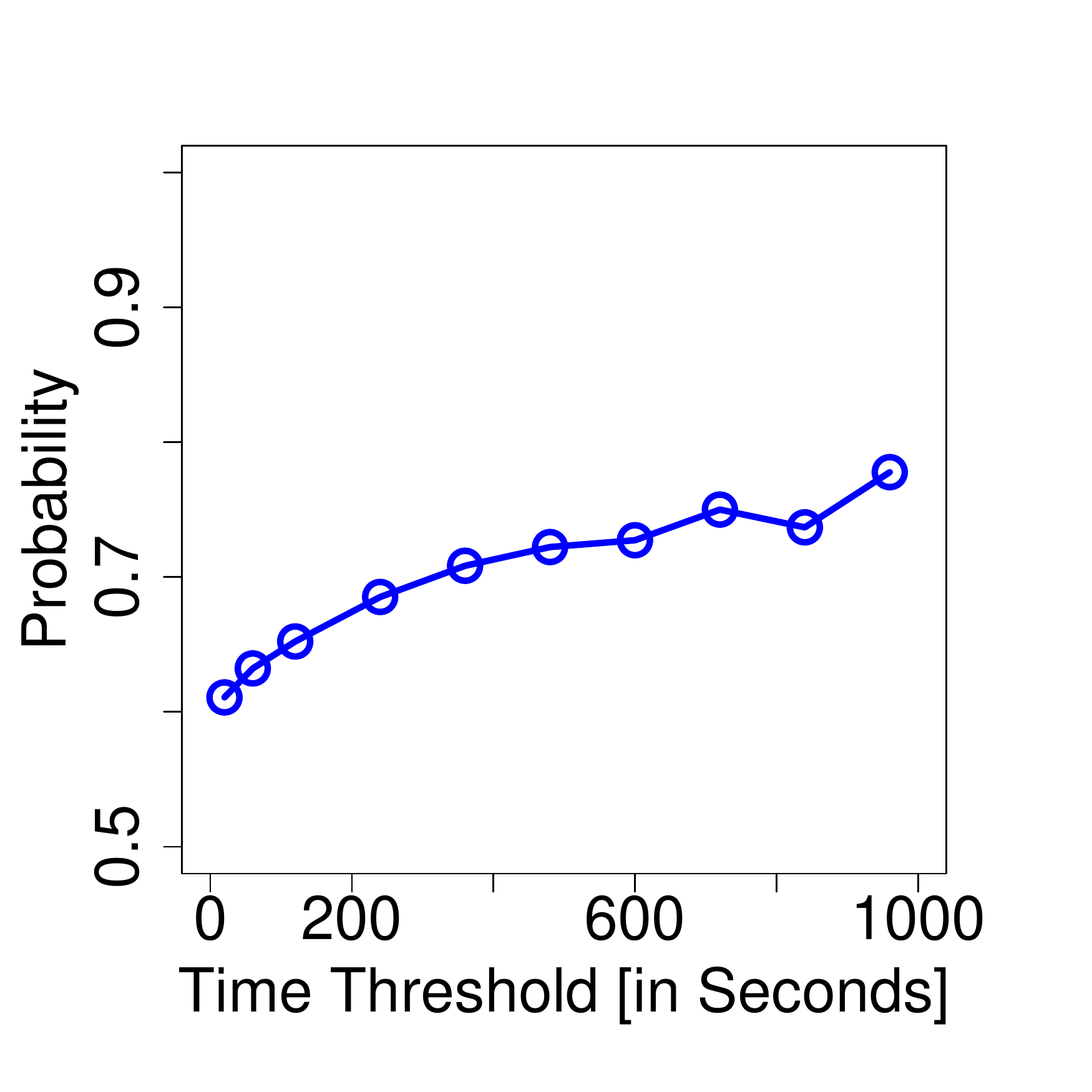}
      \end{center}
      \caption{Given a face-to-face contact between a conference participant $p$ and a presenter $q$. The $y$-axis shows the probability that participant $p$ visits the talk of presenter $q$. The $x$-axis defines here the minimum face-to-face contact duration between participants $p$ and $q$.}
\label{fig:influencefactor_f2f_with_presenter}
\end{figure}

In addition, we consider the connection between a conference participant and the presenter of a talk. We  study here, whether a participant $q$ will attend the talk of this presenter $p$, when there exists a face-to-face contact between participants $p$ and $q$. In Figure~\ref{fig:influencefactor_f2f_with_presenter}, we plot the probability to join the talk of presenter $p$, given that there exists a face-to-face contact with presenter $p$ with minimum contact duration of $t \ge 20$ seconds (20 seconds is the minimum contact duration).  We observe that the probability is $61 \%$ that participant $q$ attends the talk of presenter $p$, when there exists a face-to-face contact between participants $p$ and $q$. Note that the probability is $50.8 \%$) if there exists no face-to-face contact. When we focus more and more on stronger ties (this means all face-to-face contacts greater than a given time threshold) between these two participants we see that the probability increases almost linearly to attend the talk of presenter $p$. Here the probability to attend the talk is $77.78 \%$, if there exists a face-to-face contact with contact duration greater than $960$ seconds.

\subsubsection{Predictability of Talk Attendance using Simple Baseline Predictors}\label{sec:SimpleBaselinePredictors}

\begin{figure}[hbt]
     \begin{center}
      \includegraphics[width=1\columnwidth]{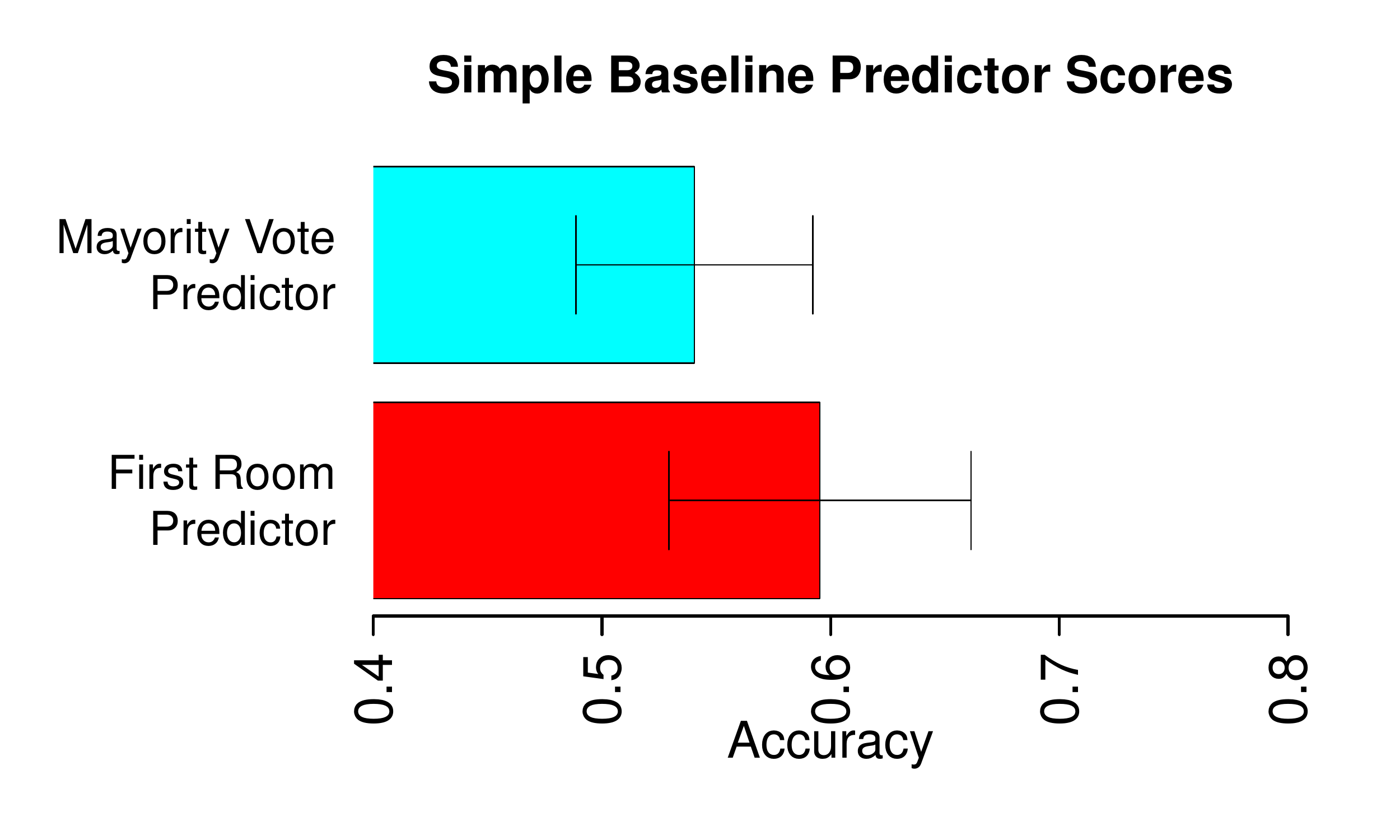}
      \end{center}
      \caption{Influence factors concerning the prediction of talks at academic conferences. The majority vote predictor predicts the next talk based on greater number of accepted papers for the corresponding tracks.}
\label{fig:influencefactor_SimplePred}
\end{figure}

Next, we analyze the prediction quality of two simple baseline models. We first predict the next talk, based on the number of the accepted papers of the corresponding tracks. This means we predict that a participant joins talk $A$ (instead of talk $B$), when the number of accepted papers of the track, which talk $A$ belongs to, is greater than the number of accepted papers for the track, which talk $B$ belongs to. We see (in Figure \ref{fig:influencefactor_SimplePred}) that the accuracy of this majority vote predictor is $54.05 \%$. Furthermore, we predict the next talk that the conference participant is going to attend, based on the room of the first talk this participant attended. Here the prediction accuracy is $59.53\%$. 
  
\subsubsection{Predictability of Talk Attendance based on User Interests}
We also investigated to what extent conference participants at HT 2011
decided for their attended talks based on the topics of the talks. This is motivated
by the general conception that, next to social interactions, personal interest in the presented topics is another
major influence factor for talk attendance decisions.

For our analysis, we assume that personal interest is reflected by previous
publications. While modeling user interest this introduces limitations with resepct to novel
upcoming topics, it is based on observable facts and thus leads to simple inference.
We downloaded all accessible publications of a user with a publication date before the beginning of the conference. From
these, we counted word occurrences into bag-of-word models.

All bag-of-word models were generated after removing stopwords, stemming word tokens
using the porter stemmer, and tf-idf weighting.
In the most simple setup, we estimate similarities between a visitor's interest
and the topic of a talk by calculating the cosine similarity between the
respective bag-of-word vectors. For each user and time slot, we predict which out
of two parallel talks is attended. The predictor itself is argmax, \ie  we
predict the talk with the higher cosine similarity to the participants interest
model.
In order to avoid further influence factors in our experiments, we evaluate content-based influence on a
core of our dataset. In this core, only those 51 out of the original 53 users are
retained, for which we were able to download at least one prior publication.

To model topics of talks, we build bag-of-word vectors directly from the presented
papers in the proceedings. For additional experiments, we also limit to bag-of-word
models derived only from abstracts or paper titles.

To find out more about the topical separation of parallel sessions with respect to
cosine similarity, we calculated and compared all talk-talk similarities inside and across
sessions in the same time-slot.

Given our observation that participants at HT 2011 changed between
sessions very infrequently, it appears that conference participants decide for attending whole sessions
rather than individual talks. 
If all talks inside the same session have a much stronger
topical relation than talks from different sessions, then every individual talk is
already a good representative for the session topic. In this case it does not matter which single talk is used to predict the attended session. Otherwise, prediction should be based on all similarities to all the talks in the session.
To find out about how well each talk is associated to its session according to the cosine measure,
we apply a cluster quality analysis. Here, clustering does not refer to the oucome
of a clustering algorithm but to the true distribution of talks over sessions.
For each pair of parallel sessions, we calculate the average silhouette value \cite{silhouette} over all talks of both
sessions. The silhouette value adopted to our task is defined as
$
silh(t) = \frac{dist(t,s_{\neg t}) - dist(t, s_t)}{max(dist(t,s_{\neg t}) , \: dist(t,
s_t))}
$
, where $t$ is a certain talk, $s_t$ the session, to which $t$ belongs, and $s_{\neg t}$
is the other session at the same time. We choose single-link cosine distance,
\ie $dist(a,b) = 1 - cos(a,b)$ where $cos$ captures the similarity to the closest non-identical talk within the respective session.

\begin{figure}[hbt]
\begin{minipage}{1\columnwidth}
      \begin{center}
      \includegraphics[width=1.01\columnwidth]{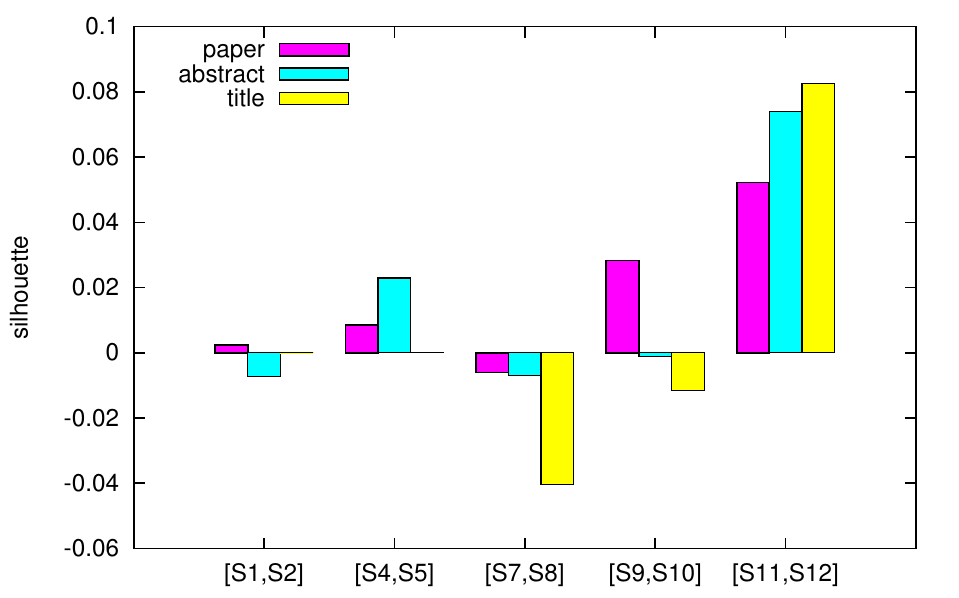}
       \end{center}
\end{minipage}
\caption{
Cluster quality of the true distribution of talks over parallel
sessions; measured by silhouette coefficient calculated using cosine distance based on different talk representations
}
\label{figSilhouette}
\end{figure}

Figure \ref{figSilhouette} compares the average silhouette values for each set of
parallel sessions.  We observe, that all values are
relatively close to zero. This means that none of
the talk representations gives a good explanation for the distribution of talks
over sessions although some talk distributions over session pairs, especially for
sessions 11 and 12, show slighlty higher silhouette values.

\begin{figure}[hbt]
\subfigure[]{
\begin{minipage}{1\columnwidth}
      \begin{center}
      \includegraphics[width=1.01\columnwidth]{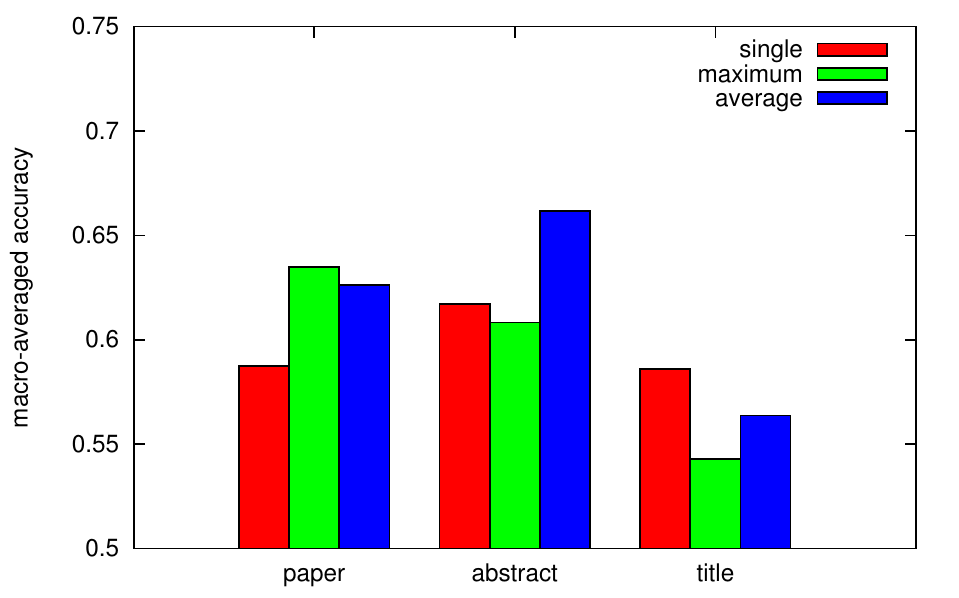}
       \end{center}
\end{minipage}
\label{figCOS_ACC}
}
\hfill
\subfigure[]{
\begin{minipage}{1\columnwidth}
      \begin{center}
      \includegraphics[width=1.01\columnwidth]{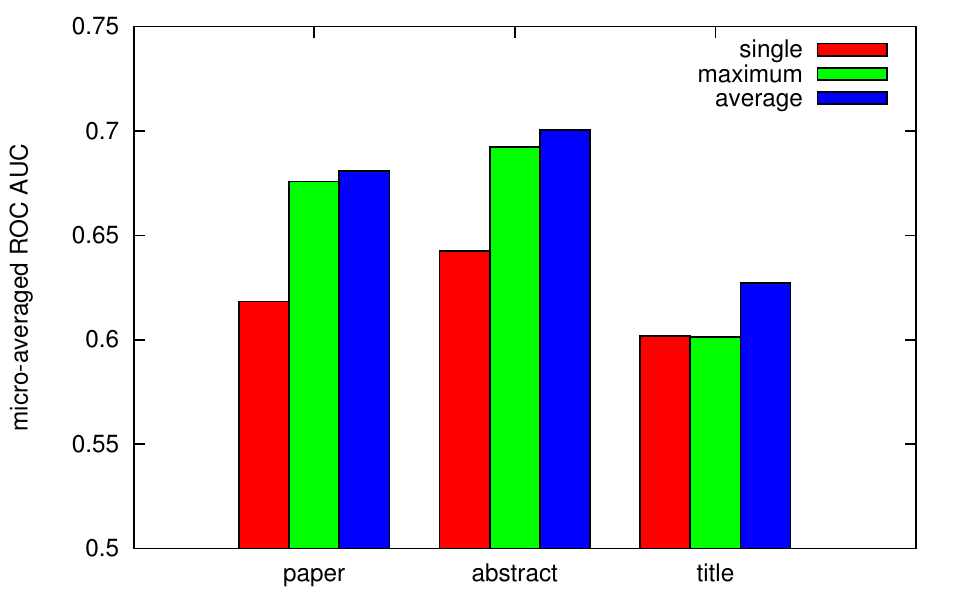}
       \end{center}
\end{minipage}
\label{figCOS_AUC}
}
 \caption{Accuracy (a) and AUC (b) values for content-based talk prediction based
 on cosine similarity depending on the talk representation and whether the decision is predicted for each talk separately or once
 for all talks of a session by either maximum or average talk similarity}
\label{figCOS_PAPER_ABSTRACT_TITLE}
\end{figure}
Motivated by the intuition that people might make their talk attendance decisions
based on one or two talks rather than on all talks of the session, and that participants do not change sessions, we also predict
talk attendance session-wise.   
We tried two options for predicting each users attended session. For the first option, we predict the session with
the higher maximum similarity of a talk in the respective session. For the second option, we choose the session
with the higher average talk similarity. Figure \ref{figCOS_ACC} depicts the accuracy values averaged over all
$337$ individual decisions of a certain participant for a certain talk. 
 Figure \ref{figCOS_AUC} further shows the area under the ROC curve results. For building up the
global micro-average ranking, we first normalized all cosine scores of each
particular person and time-slot by dividing by their sum.

As can be seen from Figure \ref{figCOS_AUC},
session-wise prediction constantly achieves superior AUC scores. While this is in accordance to our finding that only few people change between parallel
sessions, it is also although the silhouette analysis revealed that the distribution of talks over parallel
sessions is not clearly explainable by the cosine-based model. Overall, this
can be interpreted as an indication that, from the model's perspective, the decision of a
participant for one of the session is hard, and, that it is usually a
consideration of one or more of the most interesting talks.
    For the global prediction ranking measured by the AUC, the average
attendee-talk similarity constantly results in the best results. For the accuracy,
the session-wise predictors are also favorable for paper and abstract based talk
models. An exception is the title-based model. There both session-wise predictors
score below the talk-wise predictor. Title-based models, produce very sparse
bag-of-word vectors and might thus lead to low or zero scores for both parallel
talks. Yet, the talk-wise predictor achieves a mean accuracy that is only slightly
lower compared to other, less sparse models. This indicates that title words are
good topic indicators. Lower results for the session-wise predictors, especially
the maximum predictor, may be due to wrong decisions based on few overly weighted word
matches in one of the talks' titles. This might also be one reason for the low
maximum predictor accuracy on abstract-based talk models. Still, the confidence
scores of low quality predictions must be low enough to not influence the global
AUC ranking. 
Interestingly, the results with abstract-based talk models constantly compare favorably to
full-paper models in all settings. Potential explanations are the summarizing
character of abstracts and the fact that people often do not have the
opportunity to read the full paper before choosing the talk to attend.
In order to avoid the sparsity problem, we also experimented using dimensionality reduction like for example used by \cite{schuetze1998} which is
similar to \emph{Latent Semantic Indexing}\cite{lsi}. SVD is used here to map to a
denser lower dimensional vector space spanned by the $k$th largest eigenvalues of a matrix
$M M^T$. The columns of $M$ are the sums of
all context window vectors of a token. Context vectors are derived from sliding
windows around each token in the previous papers of the (tracked) participants.
However, results did not lead to a clear improvement.

Furthermore, we experimented with a feature-selection based on the pointwise
mutual information (PMI) of the two conditional probability estimates for
\emph{some author used term $t$ in his previous papers} and \emph{some author used term $t$ in the proceedings} given a term $t$. This PMI was
intended to measure cosine similarity only based on tokens which are to some
extend related to the topics of the conference. However, this also did not yield a
better interest model for explaining talk attendances. A potential cause is the
small number of presentations and the resulting low amount of data for estimating
the probabilities.

%
%

\begin{figure*}[hbt]
\subfigure[Cosine Network]{
\begin{minipage}{1\columnwidth}
      \begin{center}

\begin{tikzpicture}
[
	remember picture,
	scale=0.97, transform shape,
	person/.style={circle,draw=blue!50,fill=blue!10,thick,inner sep=0pt,minimum size=6mm},
	talk/.style={circle,draw=red!50,fill=red!20,thick,inner sep=0pt,minimum	size=6mm},
	every fit/.style={ellipse,draw,inner sep=0pt},
]

\node (p1) [person] {$p_1$};
\node (p3) [person, above=of p1] {$p_3$};
\node (p2) [person, left=of p3] {$p_2$};
\node (p4) [person, right=of p3] {$p_4$};
\node (t1) [talk, above left=1.5 and 0.25 of p2] {$t_1$};
\node (t2) [talk, right=of t1] {$t_2$};
\node (t3) [talk, above left=1.5 and 0.25 of p4] {$t_3$};
\node (t4) [talk, right=of t3] {$t_4$};
\node (src) [below=0.5 of p1] {};

\node[ellipse, draw, fit=(t1) (t2), label=above:session 1, gray] {};
\node[ellipse, draw, fit=(t3) (t4), label=above:session 2, gray] {};

\draw[->] (p1) .. controls +(left:35mm) and +(left:10mm) .. (t2);
\draw[->] (p1) .. controls +(right:35mm) and +(right:10mm) .. (t3);

\path[->]  (src) edge (p1)
	(p1)	edge[bend left=80] (t1)
		edge[bend right=80](t4)
	(p2)	edge (t1)
		edge (t2)
		edge (t3)
		edge (t4)
	(p3)	edge (t1)
		edge (t2)
		edge (t3)
		edge (t4)
	(p4)	edge (t1)
		edge (t2)
		edge (t3)
		edge (t4);

\end{tikzpicture}
       \end{center}
\end{minipage}
\label{figPrCos}
}
\hfill
\subfigure[Coffee-Break Network]{
\begin{minipage}{1\columnwidth}
      \begin{center}

\begin{tikzpicture}
[
	remember picture,
	scale=0.97, transform shape,
	person/.style={circle,draw=blue!50,fill=blue!10,thick,inner sep=0pt,minimum size=6mm},
	talk/.style={circle,draw=red!50,fill=red!20,thick,inner sep=0pt,minimum	size=6mm},
	every fit/.style={ellipse,draw,inner sep=0pt},
]
\node (cp1) [person] {$p_1$};
\node (cp3) [person, above=of cp1] {$p_3$};
\node (cp2) [person, left=of cp3] {$p_2$};
\node (cp4) [person, right=of cp3] {$p_4$};
\node (ct1) [talk, above left=1.5 and 0.25 of cp2] {$t_1$};
\node (ct2) [talk, right=of ct1] {$t_2$};
\node (ct3) [talk, above left=1.5 and 0.25 of cp4] {$t_3$};
\node (ct4) [talk, right=of ct3] {$t_4$};
\node (csrc) [below=0.5 of cp1] {};

\node[ellipse, draw, fit=(ct1) (ct2), label=above:session 1, gray] {};
\node[ellipse, draw, fit=(ct3) (ct4), label=above:session 2, gray] {};

\path[<->]  (csrc) edge[->] (cp1)
	(cp2)	edge (cp3)
		edge (cp1)
	(cp4)	edge (cp1)
		edge (ct3)
	(ct1)	edge (ct2)
		;

\end{tikzpicture}
       \end{center}
\end{minipage}
\label{figPrContact}
}
\subfigure[Presenter Network]{
\begin{minipage}{2\columnwidth}
      \begin{center}

\begin{tikzpicture}
[
	remember picture,
	scale=0.97, transform shape,
	person/.style={circle,draw=blue!50,fill=blue!10,thick,inner sep=0pt,minimum size=6mm},
	talk/.style={circle,draw=red!50,fill=red!20,thick,inner sep=0pt,minimum	size=6mm},
	every fit/.style={ellipse,draw,inner sep=0pt},
]
\node (pp1) [person] {$p_1$};
\node (pp3) [person, above=of pp1] {$p_3$};
\node (pp2) [person, left=of pp3] {$p_2$};
\node (pp4) [person, right=of pp3] {$p_4$};
\node (pt1) [talk, above left=1.5 and 0.25 of pp2] {$t_1$};
\node (pt2) [talk, right=of pt1] {$t_2$};
\node (pt3) [talk, above left=1.5 and 0.25 of pp4] {$t_3$};
\node (pt4) [talk, right=of pt3] {$t_4$};
\node (psrc) [below=0.5 of pp1] {};

\node[ellipse, draw, fit=(pt1) (pt2), label=above:session 1, gray] {};
\node[ellipse, draw, fit=(pt3) (pt4), label=above:session 2, gray] {};

\path[<->]  (psrc) edge[->] (pp1)
	(pp1)	edge[bend left=50] (pt1)
	(pp3)	edge (pt2)
	(pp4)	edge (pt3)
	(pt1)	edge[bend right] (pt3)
		edge (pt2)
		;

\end{tikzpicture}
       \end{center}
\end{minipage}
\label{figPrContact}
}
\caption{Example illustrations of the \emph{Hybrid Rooted PageRank} networks for
talk prediction. Graph (a) shows the structure of a cosine (user interests) network connecting
persons $p_1, ... , p_4$ with the presenters of the talks $t_1, ... , t_4$ for which attendance is to be predicted.
Graph (b) shows the coffee-break network linking persons for whom face-to-face contact
have been measured in the coffee-break before the talk. Graph (c) is the presenter
network, containing face-to-face contacts at any time before the talks to be
predicted. Dashed links show an example part for switching the
networks. Such links exist between all nodes with equal labels.}
\label{fig:hrpr_differentnetworks}

\begin{tikzpicture}[remember picture,overlay]
\path[<->,very thick, dashed, orange, bend left]
	(p2)	edge (cp2)
	(cp2)	edge (pp2)
	(pp2)	edge (p2);
\end{tikzpicture}

\end{figure*}
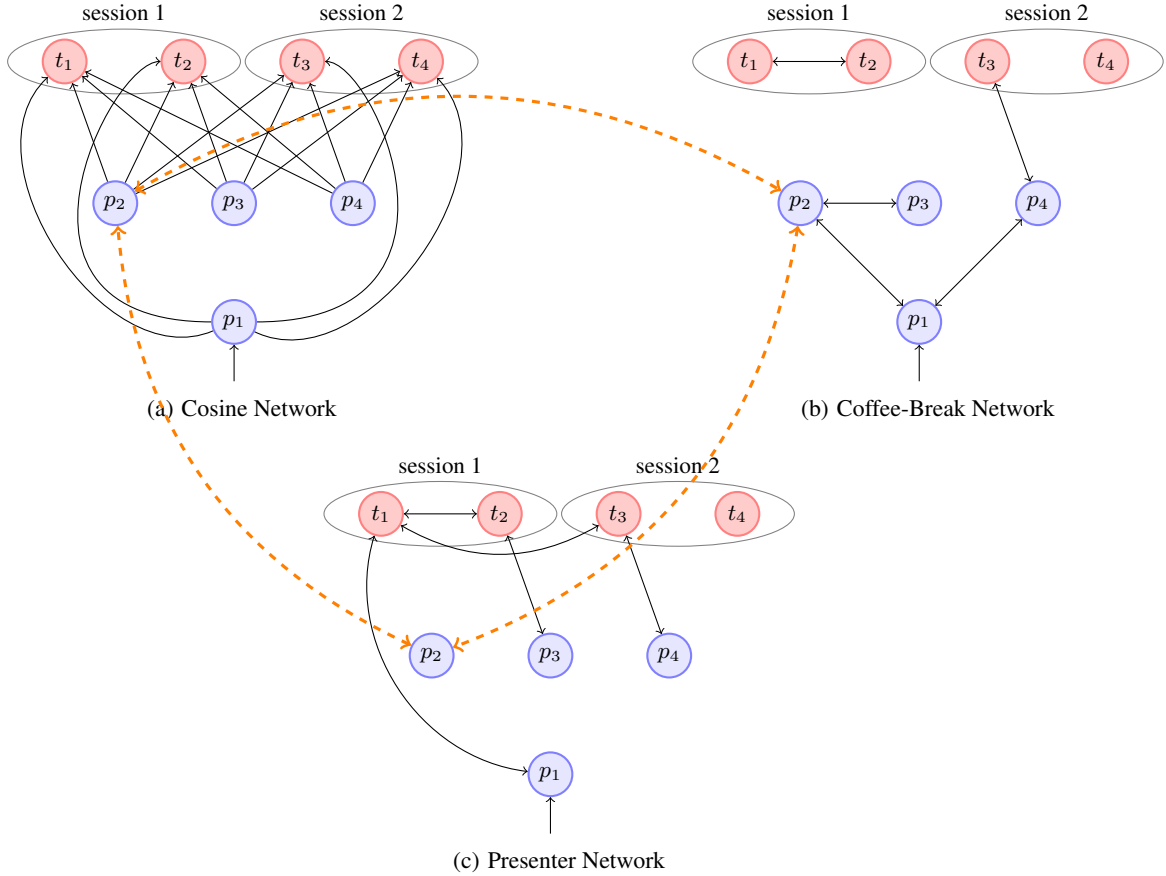

\subsubsection{Predictability of Talk Attendance Using the Hybrid Rooted PageRank Predictor}
In this subsection, we analyze the predictability of talk attendance using a combination of different networks. For this analysis, we use the \emph{Hybrid Rooted PageRank} (HRPR) algorithm (see section \ref{sec:algorithms:hrpr}) as predictor. The advantage of this algorithm is that we can analyze and compare the predictive power of different networks and combinations of these networks. Using the HRPR-algorithm, we combine the information of the paper-similarity network, the aggregated face-to-face contact network (of the coffee break before the next talk is going to start), and the presenters face-to-face contact network. The structure of these graph is illustrated in Figure \ref{fig:hrpr_differentnetworks}. Note that the hybrid rooted random walk (see Algorithm \ref{alg:hybridrandomwalk}) selects a network with respect to a given probability distribution $P=(p_1,p_2,p_3)$. In our experiments, we studied all parameter combinations with $p_1 + p_2 + p_3 = 1$ and $p_1,p_2,p_3 \in \{0.0,0.1,0.2,\ldots,1.0\}.$ Assume we want to predict, whether participant $p$ attends talk $t_1$ or talk $t_2$. The predicted talk is then given by the talk $t_i$, where $p_i$ is the presenter of talk $t_i$, $p_j$ the presenter of talk $t_j$ and $HRPR(p,p_i) > HRPR(p,p_j)$.

 We start by analyzing the predictive power for each network separately. In Figure \ref{fig:auc_differentnetworks_notgrouped} a) and b), we observe that the paper-similarity network performs best with an AUC-value of $0.630$ and an accuracy of $0.610$. These results correspond to the results of the single variant in Figure \ref{figCOS_PAPER_ABSTRACT_TITLE}). Using just the face-to-face contact network of the coffee-break does not work as well as using the paper-similarity network. Here the AUC-value is $0.596$. In contrast to the coffee break's face-to-face contact network, the presenter network contains just the links from the presenter of the next talk that is going to start. We observe that using just the presenters face-to-face contact network does not perform very well and works worse than using just the face-to-face contact network. This is because most participants do not have a face-to-face contact to a presenter before the presenter's talk starts. Hence, the presenter network is rather sparse and does not provide major predictive power on its own. The AUC-value for the presenter network is $0.474$. In this context, the observation that the face-to-face contact network works better than the presenter networks suggests that links between participants help further to improve the predictive power. 

Furthermore we analyze, whether the combination of different networks increases the predictability of talk attendance at academic conferences. In Figure~\ref{fig:auc_differentnetworks_notgrouped}, we observe that the best result can be obtained by combining the information of all networks. 
 However, the increase of predictability by combining the information of different networks is rather small, and we do not know the parameter combinations leading to the best results. The result just gives an indication that a combination can help to increase prediction quality. In our analysis, we handle the presenter network as an additional network. This gives us the possibility to weight a link between a participant and presenter separately. We observe here that the predictability could not be increased when we combine the presenter network and the face-to-face contact network. 

In Table \ref{tbl:SessionStatistics}, we observed that most participants visited all talks in one session. Furthermore, it was unlikely that a participant changed a session. Despite this observation, it is natural to assume that a participant is not interested in each talk of one session. We argue here that, in most cases, at most one or two talks of a session are the cause for attending the session. Therefore, for each network, we merge the nodes of all presenters in one session. The merged nodes thus represent the whole session. The weight vectors for in- and out-going edges are calculated as the re-normalized sum of the respective individual nodes' weight vectors. The merged network results depicted in Figure \ref{fig:auc_differentnetworks_grouped} clearly show an increase in talk prediction quality. Considering the best tested parameter combinations, the AUC score increases from $0.638$ to $0.703$ and accuracy increases from $0.617$ to $0.666$. We also observe that, for each parameter combination, the combination of all networks performs better, when we merge the presenter nodes. Unlike the model where we do not merge the presenter nodes of one session, we observe that the combination of the presenter network and the face-to-face contact network increases the prediction accuracy significantly, when we merge the presenter nodes. Considering the best parameter combinations for the presenter and face-to-face contact network, the prediction quality increases from $0.611$ to $0.68$ AUC. A further interesting point is that a minimal fraction of the face-to-face contact network or paper-similarity network increases the predictive power of the presenters face-to-face contact network from $0.61$ to $0.661$ and $0.655$. For our surprise, this trend can not be observed for the face-to-face contact network results. In addition, we observe that our presented approaches significantly outperform the simple predictors presented in Section \ref{sec:SimpleBaselinePredictors}. 

\begin{figure*}
\subfigure[Single Variant]{
\begin{minipage}{0.97\columnwidth}
      \begin{center}
      \includegraphics[width=0.97\columnwidth]{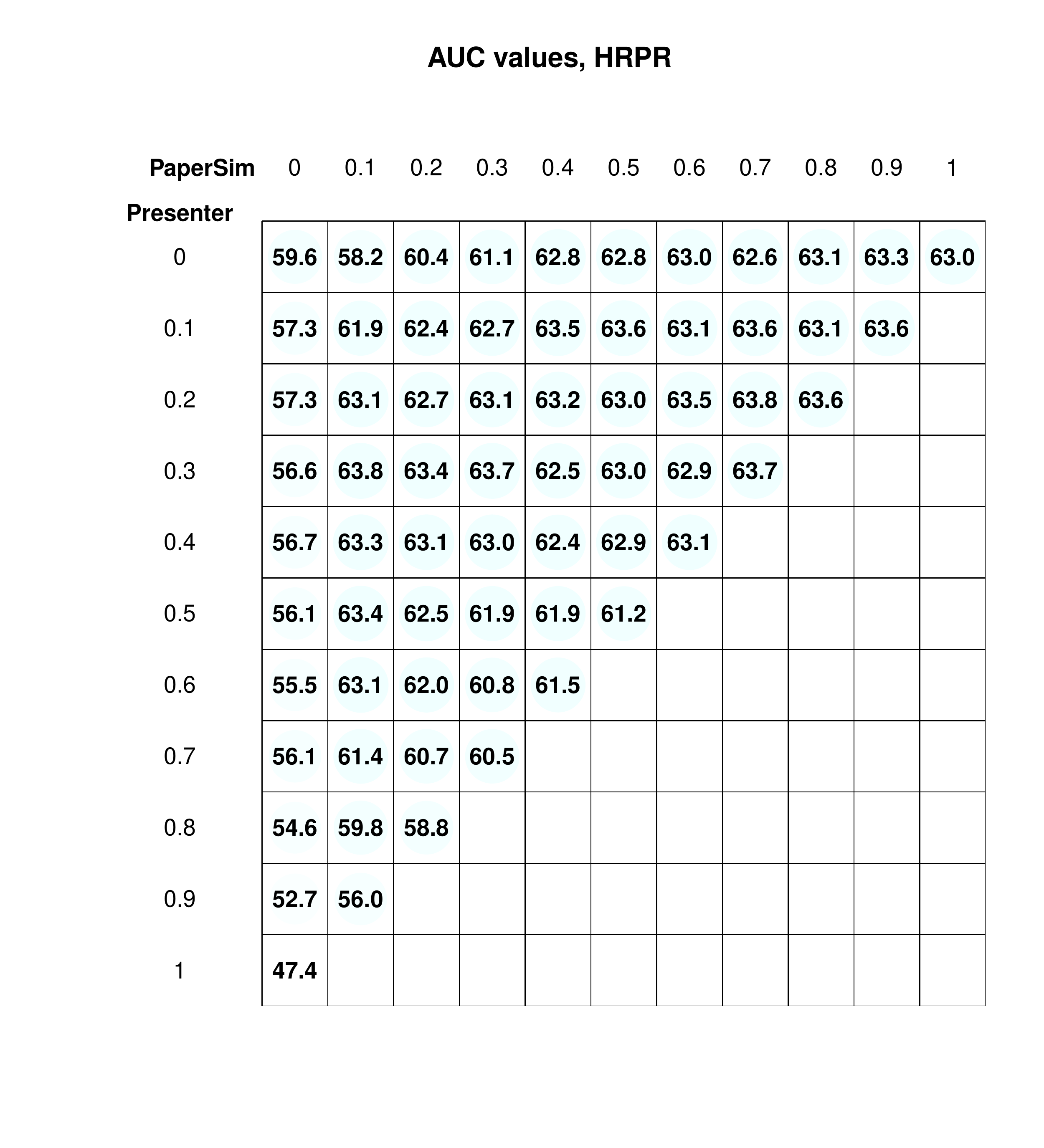}
       \end{center}
\end{minipage}
\label{fig:auc_differentnetworks_notgrouped}
}
\hfill
\subfigure[Single Variant]{
\begin{minipage}{0.97\columnwidth}
      \begin{center}
      \includegraphics[width=0.97\columnwidth]{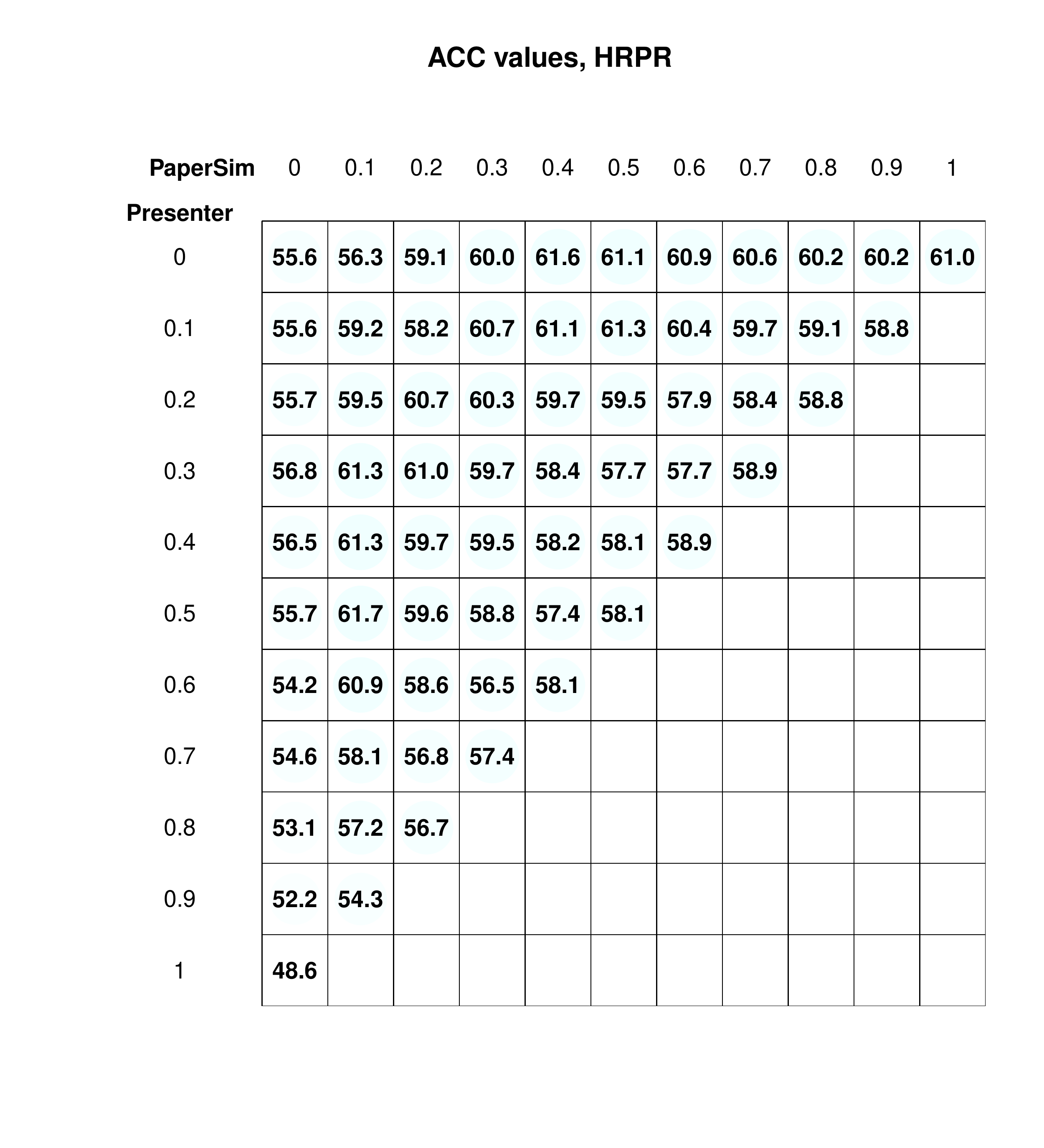}
       \end{center}
\end{minipage}
\label{fig:auc_differentnetworks_notgrouped_acc}
}
\caption{AUC and accuracy values for the talk prediction task using the \emph{Hybrid Rooted PageRank} as predictor. The $x$-axis represents the probability to choose the paper-similarity network in the random walk of $HRPR$, the $y$-axis the probability to choose the presenter network. The probability to choose the coffee-break network is then defined as 1-x-y. The $z$-axis displays the AUC-value for the defined (by the $x$ and $y$ axes) parameter combinations. In these figures we present the predictability-results, without merging the presenter nodes.}
\label{fig:auc_differentnetworks_notgrouped}
\end{figure*}

\begin{figure*}
\subfigure[Merged Variant]{
\begin{minipage}{0.97\columnwidth}
      \begin{center}
      \includegraphics[width=0.97\columnwidth]{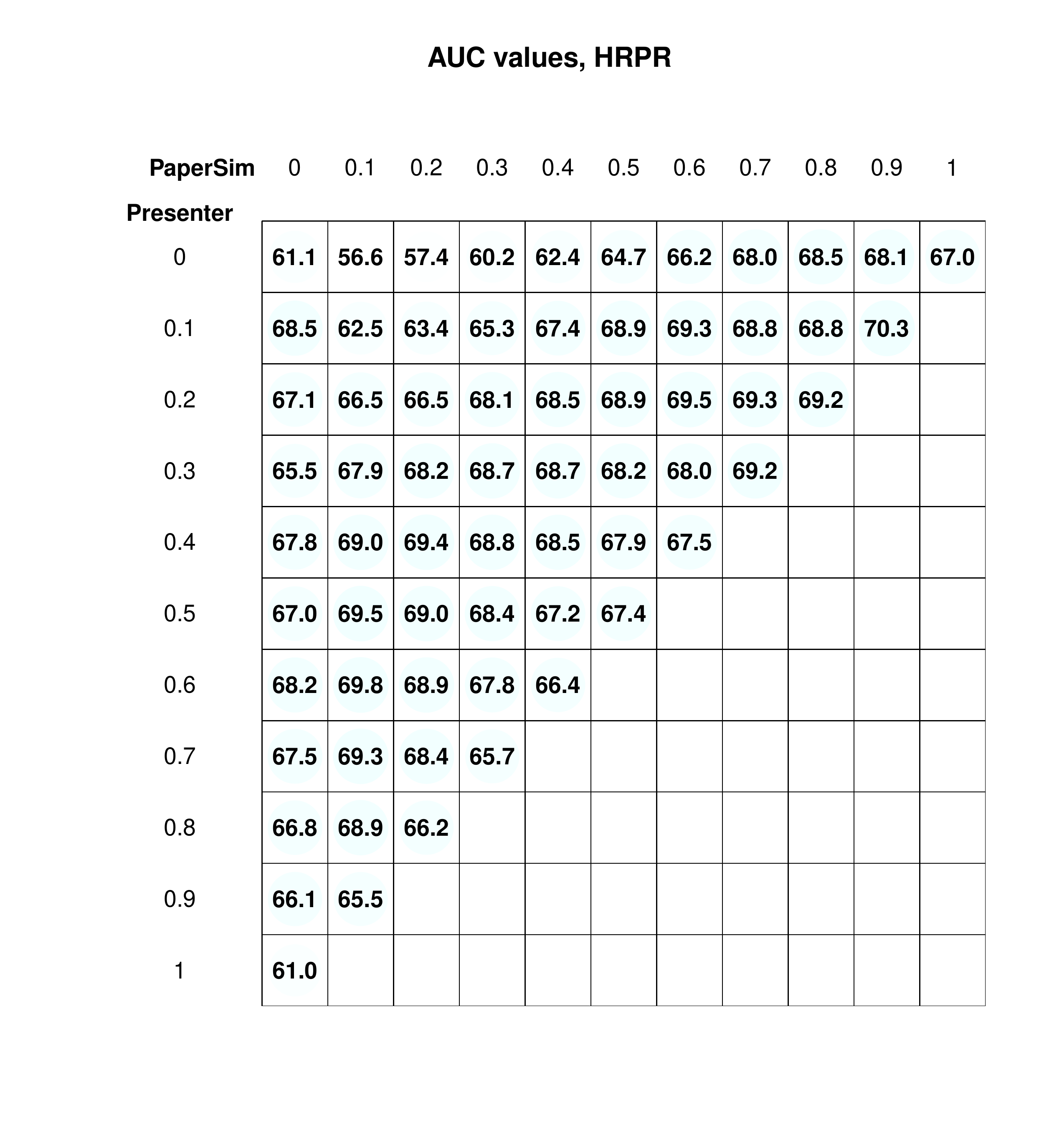}
       \end{center}
\end{minipage}
\label{fig:auc_differentnetworks_grouped}
}
\hfill
\subfigure[Merged Variant]{
\begin{minipage}{0.97\columnwidth}
      \begin{center}
      \includegraphics[width=0.97\columnwidth]{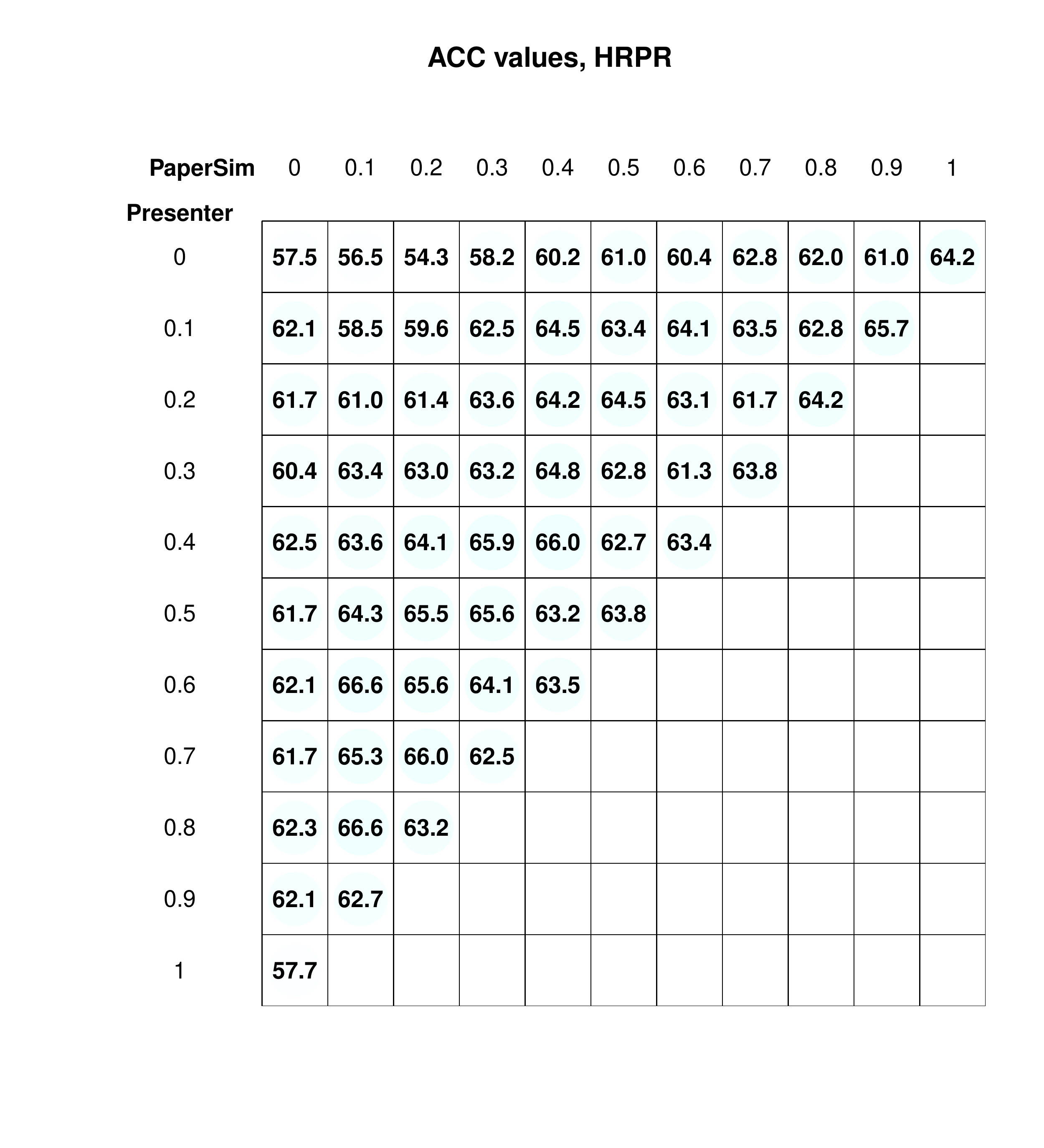}
       \end{center}
\end{minipage}
\label{fig:auc_differentnetworks_grouped_acc}
}
\caption{AUC and accuracy values for the talk prediction task using the \emph{Hybrid Rooted PageRank} as predictor. The $x$-axis represents the probability to choose the paper-similarity network in the random walk of $HRPR$, the $y$-axis the probability to choose the presenter network. The probability to choose the coffee-break network is then defined as 1-x-y. The $z$-axis displays the AUC/Accuracy-value for the defined (by the $x$ and $y$ axes) parameter combinations. In these figures we present the predictability-results, when we merge the presenter nodes.}
\label{fig:auc_differentnetworks_grouped}
\end{figure*}

\section{Conclusions}\label{sec:conclusion}

In this paper, we analyzed and discussed the predictability of talk attendance at academic conferences evaluated on real-world data. We considered different influence factors, concerning this prediction problem. 
 Specifically, we studied the influence of face-to-face contacts and user interests on the talk attendance. We showed that the probability of two participants attending the same talk is nearly random, if there exists no face-to-face contact before the talk is going to start. In this context, the probability (that two participants attend the same talk) is significantly increased if there exists a face-to-face contact in the break before the talk. Next, we analyzed the influence of user interest on talk attendance. We observed, that prediction based on user-interest alone achieves better results than prediction based solely on face-to-face contact data. Using the \emph{Hybrid rooted PageRank} we showed that a combination of different networks helps to further improve the prediction accuracy. Another important observation is that the combination of all information belonging to one session, \ie merging the presenter nodes, significantly improves prediction accuracy. This supports the theory that, in many cases, only one or two talks are the cause for a user's session attendance decision. Since it is unknown which of the talks in a session is relevant for the decision, all have to be considered.

\bibliographystyle{abbrv}

\begin{thebibliography}{10}

\bibitem{AT:05}
G.~Adomavicius and A.~Tuzhilin.
\newblock {Toward the Next Generation of Recommender Systems: A Survey of the
  State-of-the-Art and Possible Extensions}.
\newblock {\em Knowledge and Data Engineering}, 17(6), 2005.

\bibitem{ALANI09}
H.~Alani, M.~Szomszor, C.~Cattuto, W.~V. den Broeck, G.~Correndo, and
  A.~Barrat.
\newblock {Live Social Semantics}.
\newblock In {\em Intl. Semantic Web Conference}, pages 698--714, 2009.

\bibitem{ubicon-2014a}
M.~Atzmueller, M.~Becker, M.~Kibanov, C.~Scholz, S.~Doerfel, A.~Hotho, B.-E.
  Macek, F.~Mitzlaff, J.~Mueller, and G.~Stumme.
\newblock {Ubicon and its Applications for Ubiquitous Social Computing}.
\newblock {\em New Review of Hypermedia and Multimedia}, 20(1):53--77, 2014.

\bibitem{ADHMS:12}
M.~Atzmueller, S.~Doerfel, A.~Hotho, F.~Mitzlaff, and G.~Stumme.
\newblock {Face-to-Face Contacts at a Conference: Dynamics of Communities and
  Roles}.
\newblock In {\em {Modeling and Mining Ubiquitous Social Media}}, volume 7472
  of {\em LNAI}. Springer Verlag, Heidelberg, Germany, 2012.

\bibitem{CIROHIGHRES08}
A.~Barrat, C.~Cattuto, V.~Colizza, J.-F. Pinton, W.~V. den Broeck, and
  A.~Vespignani.
\newblock {High Resolution Dynamical Mapping of Social Interactions with Active
  RFID}.
\newblock {\em CoRR}, abs/0811.4170, 2008.

\bibitem{Barrat10}
A.~Barrat, C.~Cattuto, M.~Szomszor, W.~V. den Broeck, and H.~Alani.
\newblock {Social Dynamics in Conferences: Analyses of Data from the Live
  Social Semantics Application}.
\newblock In {\em International Semantic Web Conference (2)}, pages 17--33,
  2010.

\bibitem{BrinP98}
S.~Brin and L.~Page.
\newblock {The Anatomy of a Large-Scale Hypertextual Web Search Engine}.
\newblock {\em Computer Networks}, 30(1-7):107--117, 1998.

\bibitem{Cattuto:2010}
C.~Cattuto, W.~{Van den Broeck}, A.~Barrat, V.~Colizza, J.-F. Pinton, and
  A.~Vespignani.
\newblock {Dynamics of Person-to-Person Interactions from Distributed {RFID}
  Sensor Networks}.
\newblock {\em PLoS ONE}, 5(7):e11596, 07 2010.

\bibitem{SABCS:13}
{Christoph Scholz and Martin Atzmueller and Alain Barrat and Ciro Cattuto and
  Gerd Stumme}.
\newblock {New Insights and Methods For Predicting Face-To-Face Contacts}.
\newblock In {\em Proc. 7th Intl. AAAI Conference on Weblogs and Social Media},
  2013.

\bibitem{lsi}
S.~Deerwester, S.~Dumais, G.~Furnas, T.~Landauer, and R.~Harshman.
\newblock {Indexing by Latent Semantic Analysis}.
\newblock {\em Journal of the American Society for Information Science 41},
  pages 391--407, 1990.

\bibitem{eagle2009}
N.~Eagle, A.~Pentland, and D.~Lazer.
\newblock {From the Cover: Inferring Friendship Network Structure by using
  Mobile Phone Data}.
\newblock {\em Proceedings of The National Academy of Sciences},
  106:15274--15278, 2009.

\bibitem{Hanley1982}
J.~A. Hanley and B.~J. McNeil.
\newblock {The Meaning and Use of the Area under a Receiver Operating
  Characteristic (ROC) Curve}.
\newblock {\em Radiology}, 143(1):29--36, Apr. 1982.

\bibitem{Hui2005}
P.~Hui, A.~Chaintreau, J.~Scott, R.~Gass, J.~Crowcroft, and C.~Diot.
\newblock {Pocket Switched Networks and Human Mobility in Conference
  Environments}.
\newblock In {\em Proceedings of the 2005 ACM SIGCOMM workshop on
  Delay-tolerant networking}, WDTN '05, pages 244--251, New York, NY, USA,
  2005. ACM.

\bibitem{Isella:2011}
L.~Isella, M.~Romano, A.~Barrat, C.~Cattuto, V.~Colizza, W.~{Van den Broeck},
  F.~Gesualdo, E.~Pandolfi, L.~Rav\`a, C.~Rizzo, and A.~Tozzi.
\newblock {Close Encounters in a Pediatric Ward: Measuring Face-to-Face
  Proximity and Mixing Patterns with Wearable Sensors}.
\newblock {\em PLoS ONE}, 6:e17144, 2011.

\bibitem{DBLP:journals/corr/abs-1006-1260}
L.~Isella, J.~Stehl\'e, A.~Barrat, C.~Cattuto, J.-F. Pinton, and W.~V.~D.
  Broeck.
\newblock {What's in a Crowd? Analysis of Face-to-Face Behavioral Networks}.
\newblock {\em Journal of Theoretical Biology}, 271:166--180, 2011.

\bibitem{joachimsRankSvm}
T.~Joachims.
\newblock Optimizing search engines using clickthrough data.
\newblock In {\em Proceedings of the Eighth ACM SIGKDD International Conference
  on Knowledge Discovery and Data Mining}, KDD '02, pages 133--142, New York,
  NY, USA, 2002. ACM.

\bibitem{lee2012}
D.~Lee and P.~Brusilovsky.
\newblock {Exploring Social Approach to Recommend Talks at Research
  Conferences}.
\newblock In {\em COLLABORATECOM 2012 - 8th IEEE International Conference on
  Collaborative Computing: Networking, Applications and Worksharing}, Oct.
  2012.

\bibitem{Kleinberg2003}
D.~Liben-Nowell and J.~M. Kleinberg.
\newblock {The Link Prediction Problem for Social Networks}.
\newblock In {\em CIKM}, pages 556--559, 2003.

\bibitem{MSAS:12}
B.-E. Macek, C.~Scholz, M.~Atzmueller, and G.~Stumme.
\newblock {Anatomy of a Conference}.
\newblock In {\em Proc. 23rd ACM Conference on Hypertext and Social Media},
  pages 245--254, New York, NY, USA, 2012. ACM Press.

\bibitem{WCI:07}
M.~Meriac, A.~Fiedler, A.~Hohendorf, J.~Reinhardt, M.~Starostik, and J.~Mohnke.
\newblock {Localization Techniques for a Mobile Museum Information System}.
\newblock In {\em {Proceedings of WCI}}, 2007.

\bibitem{minkov2010}
E.~Minkov, B.~Charrow, J.~Ledlie, S.~Teller, and T.~Jaakkola.
\newblock {Collaborative Future Event Recommendation}.
\newblock In {\em Proceedings of the 19th ACM International Conference on
  Information and Knowledge Management}, CIKM '10, pages 819--828, New York,
  NY, USA, 2010. ACM.

\bibitem{pham2012}
M.~C. Pham, D.~Kovachev, Y.~Cao, G.~M. Mbogos, and R.~Klamma.
\newblock {Enhancing Academic Event Participation with Context-aware and Social
  Recommendations}.
\newblock In {\em Proceedings of the 2012 International Conference on Advances
  in Social Networks Analysis and Mining (ASONAM 2012)}, ASONAM '12, pages
  464--471, Washington, DC, USA, 2012. IEEE Computer Society.

\bibitem{silhouette}
P.~Rousseeuw.
\newblock {Silhouettes: A Graphical Aid to the Interpretation and Validation of
  Cluster Analysis}.
\newblock {\em Journal of Computational and Appl. Mathematics}, 20(1):53--65,
  1987.

\bibitem{SDAHS:11}
C.~Scholz, S.~Doerfel, M.~Atzmueller, A.~Hotho, and G.~Stumme.
\newblock {Resource-Aware On-Line RFID Localization Using Proximity Data}.
\newblock In {\em Proc. ECML/PKDD 2011}, 2011.

\bibitem{schuetze1998}
H.~Schütze.
\newblock {Automatic Word Sense Discrimination}.
\newblock {\em Computational Linguistics}, 24(1):97--123, 1998.

\bibitem{Stehl2011}
J.~Stehl\'e, N.~Voirin, A.~Barrat, C.~Cattuto, L.~Isella, J.-F. Pinton,
  M.~Quaggiotto, W.~Van~den Broeck, C.~R\'egis, B.~Lina, and P.~Vanhems.
\newblock {High-Resolution Measurements of Face-to-Face Contact Patterns in a
  Primary School}.
\newblock {\em PLoS ONE}, 6(8):e23176, 08 2011.

\bibitem{PorterStemmerAlgorithm}
C.~van Rijsbergen, S.~Robertson, and M.~Porter.
\newblock {New Models in Probabilistic Information Retrieval}.
\newblock 1980.

\bibitem{WBP:10}
C.~Wongchokprasitti, P.~Brusilovsky, and D.~Para.
\newblock {Conference Navigator 2.0: Community-Based Recommendation for
  Academic Conferences}.
\newblock In {\em Proc. Workshop Social Recommender Systems, IUI'10}, 2010.

\bibitem{XCWCZYWZ:11}
B.~Xu, A.~Chin, H.~Wang, L.~Chang, K.~Zhang, F.~Yin, H.~Wang, and L.~Zhang.
\newblock {Physical Proximity and Online User Behavior in an Indoor Mobile
  Social Networking Application}.
\newblock In {\em Proc. 4th IEEE Intl. Conf. on Cyber, Physical and Social
  Computing (CPSCom 2011)}, 2011.

\bibitem{Zuo:2012}
X.~Zuo, A.~Chin, X.~Fan, B.~Xu, D.~Hong, Y.~Wang, and X.~Wang.
\newblock {Connecting People at a Conference: A Study of Influence Between
  Offline and Online Using a Mobile Social Application}.
\newblock In {\em Connecting People at a Conference}, 2012.

\end{thebibliography}

\end{document}